\documentclass[12pt]{article}
\usepackage{amsmath,epsfig,graphicx} 
\input{epsf} 
\def\ltap{\raisebox{-.6ex}{\rlap{$\,\sim\,$}} \raisebox{.4ex}{$\,<\,$}} 
\def\gtap{\raisebox{-.6ex}{\rlap{$\,\sim\,$}} \raisebox{.4ex}{$\,>\,$}}

\newcommand\as{\alpha_{\mathrm{S}}} 
\newcommand\f[2]{\frac{#1}{#2}} 
\def\beq{\begin{equation}} 
\def\eeq{\end{equation}} 
\def\beeq{\begin{eqnarray}} 
\def\eeeq{\end{eqnarray}} 
 
\def\to{\rightarrow}
\def\ito{\leftarrow} 
\def\nn{\nonumber} 
 
\def\tL{{\tilde L}}
 
\def\qt{q_T}

\begin{document} 
\begin{titlepage}
\begin{flushright}
KA-TP-23-2008\\
SFB/CPP-08-69
\end{flushright}
\renewcommand{\thefootnote}{\fnsymbol{footnote}}
\vspace*{2cm}

\begin{center}
{\Large \bf Transverse-momentum resummation:}
\vskip 0.2cm
{\Large \bf a perturbative study of $Z$ production at the Tevatron}

\end{center}

\par \vspace{2mm}
\begin{center}
{\bf Giuseppe Bozzi${}^{(a)}$,
Stefano Catani${}^{(b)}$,}\\ 
\vskip .2cm
{\bf Giancarlo Ferrera${}^{(b)}$,
Daniel de Florian${}^{(c)}$
and
Massimiliano Grazzini${}^{(b)}$}\\

\vspace{5mm}

${}^{(a)}$Institut f\"ur Theoretische Physik, Universit\"at Karlsruhe\\P.O.Box 6980, D-76128 Karlsruhe, Germany\\

%

${}^{(b)}$INFN, Sezione di Firenze and
Dipartimento di Fisica, Universit\`a di Firenze,\\
I-50019 Sesto Fiorentino, Florence, Italy\\


${}^{(c)}$Departamento de F\'\i sica, FCEYN, Universidad de Buenos Aires,\\
(1428) Pabell\'on 1 Ciudad Universitaria, Capital Federal, Argentina\\

\vspace{5mm}

\end{center}

\par \vspace{2mm}
\begin{center} {\large \bf Abstract} \end{center}
\begin{quote}
\pretolerance 10000

We consider transverse-momentum ($q_T$)
resummation for Drell--Yan lepton pair production in
hadron collisions. At small values of $q_T$,
the logarithmically-enhanced QCD
contributions are resummed
up to next-to-leading logarithmic accuracy. At intermediate and large values
of $q_T$,
resummation is consistently combined with the fixed-order perturbative result.
We present numerical results for $e^+e^-$ pairs from the decay of $Z$ bosons 
produced at Tevatron energies.
We perform a detailed study of the scale dependence of the results to estimate
the corresponding perturbative uncertainty.
We comment on the comparison with the available Tevatron data.

\end{quote}

\vspace*{\fill}
\begin{flushleft}
December 2008

\end{flushleft}
\end{titlepage}

\setcounter{footnote}{1}
\renewcommand{\thefootnote}{\fnsymbol{footnote}}

\section{Introduction}

The production of $W$ and $Z$ bosons in hadron collisions
is important for physics studies within and beyond the Standard Model (SM).
The large production rates and clean experimental signatures
make these processes standard candles
for calibration purposes. At the LHC, $W$ and $Z$ boson production has
even been proposed as a luminosity monitor.
In the search for new physics, an excess of di-lepton events with large 
invariant mass or missing energy may signal the production of
new gauge bosons or of SUSY
particles. 

Because of
the above reasons, it is essential to have accurate
theoretical predictions for the vector-boson production cross sections
and distributions. Theoretical predictions  with high precision demand
detailed computations of radiative corrections.
The QCD corrections to the total cross section \cite{Hamberg:1990np}
and to the rapidity distribution of the vector boson \cite{Anastasiou:2003ds}
are known up to the next-to-next-to-leading order (NNLO) in the strong coupling
$\as$. The fully exclusive NNLO calculation, including the leptonic decay 
of the vector boson, has been completed more recently \cite{Melnikov:2006di}.
As for electroweak effects, full ${\cal O}(\alpha)$ corrections have been computed
for both $W$ \cite{ewW} and $Z$ production~\cite{ewZ}.

Among the various
distributions, an important role is played by the transverse-momentum ($q_T$) spectrum of the vector boson.
In the case of $W$ production, the uncertainty in the shape
of the $q_T$ spectrum directly affects the measurement of the $W$ mass.
A good understanding of the $q_T$ spectrum of the $Z$ boson gives important
information on the production mechanism of the $W$ boson.

In the region where $q_T\sim m_V$, $m_V$ being the mass of the vector boson, the
QCD perturbative series
is controlled by a small expansion parameter, $\as(m_V)$,
and fixed-order calculations are theoretically justified.
In this region, the QCD radiative corrections are known up to next-to-leading
order (NLO) \cite{Ellis:1981hk,Arnold:1988dp,Gonsalves:1989ar}.
In the small $q_T$ region ($q_T\ll m_V$), the convergence of the fixed-order
expansion is spoiled, since the coefficients of the perturbative series are
enhanced by powers of large logarithmic terms, $\as^n\ln^m (m^2_V/q_T^2)$.
To obtain reliable predictions, these terms have to be resummed to all
perturbative orders.

The method to systematically perform all-order resummation of
classes of logarithmically-enhanced terms at small $q_T$ is known
\cite{Dokshitzer:hw}--\cite{Catani:2000vq}.
The resummed and fixed-order 
procedures at small and large values of 
$q_T$ can then be matched at intermediate values of $q_T$,
to obtain QCD predictions for the entire range of transverse momenta.
Phenomenological studies of the vector-boson
$q_T$ distribution have been performed by combining resummed and fixed-order
perturbation theory 
at various levels
of accuracy \cite{Arnold:1990yk}--\cite{Berge:2005rv}.

In Refs.~\cite{Catani:2000vq,Bozzi:2005wk} we have proposed a method to 
perform transverse-momentum
resummation that introduces some novel features.
The resummed distribution
is factorized in terms of a universal transverse-momentum form factor
and a single process-dependent hard function.
In the small-$q_T$ region, the logarithmic terms of the form factor are 
systematically resummed in exponential form by working in impact-parameter 
and Mellin-moment space. 
A constraint of perturbative unitarity is imposed on the resummed terms,
to the purpose of reducing the effect of unjustified higher-order 
contributions at large values
of $q_T$ and, especially, at intermediate values of $q_T$.
This constraint decreases the uncertainty in the matching procedure 
of the resummed and fixed-order contributions.
The method has so far been applied to SM Higgs boson 
production \cite{Bozzi:2003jy, Bozzi:2005wk, Bozzi:2007pn},
$WW$ \cite{Grazzini:2005vw} and $ZZ$ \cite{Frederix:2008vb} production, and
slepton pair production \cite{Bozzi:2006fw}.
Related methods on transverse-momentum resummation have been applied to the
transversely-polarized Drell--Yan (DY) process \cite{Jiro} and
longitudinally-polarized semi-inclusive deep-inelastic scattering (SIDIS) 
\cite{Koike:2006fn}.

The explicit form of the universal
form factor is known up to next-to-leading logarithmic (NLL) 
\cite{Kodaira:1981nh,Catani:vd}
and next-to-next-to-leading
logarithmic (NNLL) \cite{Davies:1984hs,Davies:1984sp,deFlorian:2000pr} level.
The general form of the process dependent
hard function is known up to
the first relative order in $\as$ \cite{deFlorian:2000pr}.
The hard function has been computed up to the
second relative order in $\as$ only in the case of SM Higgs boson 
production \cite{Catani:2007vq}.

In the present paper we concentrate on DY lepton pair production,
and we apply the resummation formalism of Ref.~\cite{Bozzi:2005wk}
to the production of $Z$ bosons at Tevatron energies.
In this 
work, we limit ourself to presenting results up to NLL accuracy.
We perform a detailed study of the scale dependence of our results
and we provide an estimate of the corresponding
perturbative uncertainty. We also comment on the
comparison with the available Tevatron data.
In this way, we set the stage for a forthcoming NNLL analysis, which will be
possible once the computation of the
hard function up to second order is completed. 

The paper is organized as follows. In Sect.~\ref{sec:theory} we briefly review 
our resummation formalism and we comment on its application
to vector boson production in hadron collisions.
In Sect.~\ref{sec:results} we present numerical results on $Z$ boson production
at the Tevatron. The fixed-order and resummed predictions are discussed and
compared with the data
in Sect.~\ref{sec:fo} and 
Sect.~\ref{sec:resu}, respectively.
Our results are summarized 
in Sect.~\ref{sec:summa}. 

\section{Transverse-momentum resummation}
\label{sec:theory}

The resummation formalism that we use in this paper is 
discussed in detail in Ref.~\cite{Bozzi:2005wk}. The formalism 
can be applied to a generic process in which a high-mass system
of non strongly-interacting particles  
is produced in hadron--hadron collisions. In this section we briefly recall the
main points of the formalism, by considering the specific case of the production
of a vector boson $V$ ($V=W^+,W^-,Z/\gamma^*$) that subsequently decays in a
lepton pair of invariant mass $M$.

The transverse-momentum differential cross section for this process is
written as
\begin{equation}
\label{dcross}
\f{d\sigma_V}{d q_T^2}(q_T,M,s)= \sum_{a,b}
\int_0^1 dx_1 \,\int_0^1 dx_2 \,f_{a/h_1}(x_1,\mu_F^2)
\,f_{b/h_2}(x_2,\mu_F^2) \;
\f{d{\hat \sigma}_{V ab}}{d q_T^2}(q_T, M,{\hat s};
\as(\mu_R^2),\mu_R^2,\mu_F^2) 
\;\;,
\end{equation}
where $f_{a/h}(x,\mu_F^2)$ ($a=q,{\bar q},g$) are the parton densities of 
the colliding hadrons ($h_1$ and $h_2$) at the factorization scale $\mu_F$,
$d{\hat \sigma}_{V ab}/d q_T^2$ are the
partonic cross sections, $s$ (${\hat s}=x_1x_2s$) is the hadronic 
(partonic) centre--of--mass
energy, and $\mu_R$ is the renormalization scale.

The resummation is performed at the level of the partonic cross section, 
which is first decomposed as follows:
\begin{equation}
\label{resplusfin}
\f{d{\hat \sigma}_{V\,ab}}{dq_T^2}=
\f{d{\hat \sigma}_{V\,ab}^{(\rm res.)}}{dq_T^2}
+\f{d{\hat \sigma}_{V\,ab}^{(\rm fin.)}}{dq_T^2}\, .
\end{equation}
The `resummed' component,
$d{\hat \sigma}^{({\rm res.})}_{V\, ab}$, of the cross section
contains all the logarithmically-enhanced contributions at small $q_T$,
and it has to be evaluated by resumming them to all orders in $\as$.
The `finite' component, $d{\hat \sigma}^{({\rm fin.})}_{V\, ab}$,
is free of such contributions, and it 
can thus be evaluated at fixed order in perturbation theory.

The resummation procedure of the logarithmic terms is carried out in the
impact-parameter space. The resummed component of the partonic cross section 
is then obtained by performing the inverse Fourier (Bessel) transformation with
respect to the impact parameter $b$. We write:
\begin{equation}
\label{resum}
\f{d{\hat \sigma}_{V \,ab}^{(\rm res.)}}{dq_T^2}(q_T,M,{\hat s};
\as(\mu_R^2),\mu_R^2,\mu_F^2) 
=\f{M^2}{\hat s} \;
\int_0^\infty db \; \f{b}{2} \;J_0(b q_T) 
\;{\cal W}_{ab}^{V}(b,M,{\hat s};\as(\mu_R^2),\mu_R^2,\mu_F^2) \;,
\end{equation}
where $J_0(x)$ is the $0$th-order Bessel function. 
The factor ${\cal W}$ embodies the all-order dependence on the large logarithms 
$\ln (M^2b^2)$ at large $b$, which correspond to the $q_T$-space terms
$\ln (M^2/q_T^2)$ at small $q_T$.
By considering the $N$-moments ${\cal W}_N$ of ${\cal W}$ 
with respect to the variable $z=M^2/{\hat s}$ at fixed $M$,
the resummation structure of ${\cal W}_{ab, \,N}^V$ can be factorized and
organized in exponential form\footnote{Here, to simplify the notation, 
flavour indices are understood.
In other words, we limit ourselves to discussing the flavour non-singlet 
contribution. A complete discussion
of the exponentiation structure in the general case can be found 
in Appendix A of Ref.~\cite{Bozzi:2005wk}.}:
\begin{align}
\label{wtilde}
{\cal W}_{N}^{V}(b,M;\as(\mu_R^2),\mu_R^2,\mu_F^2)
&={\cal H}_{N}^{V}\left(M, 
\as(\mu_R^2);M^2/\mu^2_R,M^2/\mu^2_F,M^2/Q^2
\right) \nonumber \\
&\times \exp\{{\cal G}_{N}(\as(\mu^2_R),L;M^2/\mu^2_R,M^2/Q^2
)\}
\;\;,
\end{align}
were we have defined the logarithmic expansion parameter $L$, 
\begin{equation}
\label{logpar}
L\equiv \ln \f{Q^2 b^2}{b_0^2} \;\;,
\end{equation}
and the coefficient $b_0=2e^{-\gamma_E}$ ($\gamma_E=0.5772...$ 
is the Euler number) has a kinematical origin.
The function ${\cal H}_N^{V}$ does not depend on the impact parameter $b$, and 
it includes all the perturbative
terms that behave as constants in the limit $b\to\infty$. 
The function ${\cal G}_N$ includes the complete dependence on $b$ and, 
in particular, it contains all the terms that order-by-order in $\as$ 
are logarithmically divergent as $b\to\infty$.

This separation (actually, factorization) between finite and divergent (or
logarithmically-en\-hanced) terms involves some degree of arbitrariness.
The arbitrariness is parametrized by the introduction of the 
resummation scale $Q$ \cite{Bozzi:2005wk}, which sets the scale of the 
expansion parameter $L$ in Eq.~(\ref{logpar}).  
Although the resummation factor ${\cal W}_{N}^{V}$ does not depend on $Q$
when evaluated at each fixed order in $\as$, its explicit dependence on $Q$ 
appears when ${\cal W}_{N}^{V}$ (and, more precisely, ${\cal G}_N$)
is computed by truncation of the resummed expression at some level of
logarithmic accuracy (see Eq.~(\ref{exponent}) below). The resummation scale 
$Q$ has to be chosen of the order of the hard scale $M$; 
variations of $Q$ around $M$ can then be used to estimate the effect
of yet uncalculated higher-order
logarithmic contributions.

All the large logarithmic terms $\as^nL^m$, with $1 \leq m \leq 2n$, are
included in the form factor $\exp\{{\cal G}_N\}$ on the right-hand side of 
Eq.~(\ref{wtilde}). More precisely, all the logarithmic contributions to 
${\cal G}_N$ with $n+2 \leq m \leq 2n$ are vanishing. Therefore, the exponent
${\cal G}_N$ can be organized in classes of logarithmic contributions that
can systematically be expanded in powers of $\as=\as(\mu^2_R)$, 
at fixed value of $\lambda=\as L$.
The logarithmic expansion of ${\cal G}_N$ reads
\begin{align}
\label{exponent}
{\cal G}_{N}(\as, L;M^2/\mu^2_R,M^2/Q^2)&=L \;g^{(1)}(\as L)+g_N^{(2)}(\as L;M^2/\mu_R^2,M^2/Q^2)\nn\\
&+\f{\as}{\pi} g_N^{(3)}(\as L,M^2/\mu_R^2,M^2/Q^2)+\dots
\end{align}
where the term $L\, g^{(1)}$ collects the leading logarithmic (LL) 
contributions, the function $g_N^{(2)}$ includes
the NLL contributions, $g_N^{(3)}$ controls the NNLL terms and so forth.

An important feature of the resummation formalism \cite{Bozzi:2005wk} 
is that the form factor $\exp\{{\cal G}_N\}$ is process independent
(and independent of the factorization scale, as well).
In other words, the functions $g_N^{(i)}$ are universal: 
they only depend on the flavour of the partons
that contribute to the cross section in each specific partonic channel.
More precisely, each function $g_N^{(i)}$ has a known functional form
that is completely specified \cite{Collins:1981uk,Collins:1984kg}
in terms of few perturbatively-computable (and process independent)
coefficients and of the customary parton anomalous dimensions 
$\gamma_{ab, N}(\as)$. These perturbative coefficients, which are
flavour-dependent, are usually denoted by 
$A_a^{(n)},  B_a^{(n)}$ and $C_{ab, N}^{(n)}$.
The explicit expressions of $g_N^{(i)}$ up to $i=3$ can be found in 
Ref.~\cite{Bozzi:2005wk}. In the case of DY production (and in any production
process that occurs through $q{\bar q}$ annihilation at the Born level),
the LL function $g^{(1)}$ depends on the coefficient $A_q^{(1)}=C_F$, 
the NLL function $g_N^{(2)}$
also depends on ${B}_q^{(1)}$ 
and $A_q^{(2)}$ \cite{Kodaira:1981nh},
and the NNLL function $g_N^{(3)}$ also depends on 
$C_{qa, N}^{(1)}$ $(a=q,g)$ \cite{Davies:1984hs}, 
${B}_q^{(2)}$ \cite{Davies:1984sp, deFlorian:2000pr} and  
$A_q^{(3)}$.
All these coefficients are known, with the sole exception of $A_q^{(3)}$.
It is usually assumed that the value of $A_q^{(3)}$
is the same as the one \cite{Vogt:2000ci,Moch:2004pa}
that appears in resummed calculations of soft-gluon contributions near
partonic threshold.

To compute the resummed component of the partonic cross section, 
Eq.~(\ref{wtilde}) has to be inserted in the right-hand side 
of Eq.~(\ref{resum}). Using the expression in Eq.~(\ref{exponent}),
the resummation of the large logarithmic contributions in 
$\exp\{{\cal G}_N\}$ affects 
not only the small-$q_T$ region ($q_T\ll M$), 
but also the large-$q_T$ region ($q_T\sim M$).
This can easily be understood by observing that the 
logarithmic expansion parameter $L$ is divergent when $b\to 0$.
To avoid the introduction of large and 
unjustified higher-order contributions in 
the small-$b$ (or, equivalently, large-$q_T$) region,
the logarithmic variable $L$ in Eq.~(\ref{logpar}) is replaced by the variable
$\tL$ \cite{Bozzi:2005wk}:
\begin{equation}
\label{ltilde}
L\to\tL~~,~~~~~~~~~~\tL\equiv \ln \left(\f{Q^2 b^2}{b_0^2}+1\right)\, .
\end{equation}
The variables $L$ and $\tL$ are equivalent (to arbitrary logarithmic accuracy)
when $Qb\gg 1$, but they lead to a different behaviour
of the form factor at small values of $b$ (i.e. large values of $q_T$).
In fact, when $Qb\ll 1$ we have $\tL\to 0$ and $\exp\{{\cal G}_N\}\to 1$.
The replacement\footnote{We observe that the divergent behaviour of 
the logarithmic parameter $\ln (M^2 b^2)$ when $b \to 0$ 
can be removed 
by a generic replacement of type $\ln (M^2 b^2) \to L_c=\ln (M^2 b^2 + c)$, 
where $c$ is some positive constant of order unity. The effect of the constant 
$c$ can be rewritten as $L_c = {\tL}_c + \ln c$, where
${\tL}_c=\ln (M^2 b^2/c + 1)$.
Using the variable $L_c$ as argument of the form factor $\exp\{{\cal G}_N\}$
would spoil the constraint of perturbative
unitarity, since the term $\ln c$ in $L_c$ does not vanish at $b=0$.
We are interested to maintain this constraint and, therefore, we have to remove
the term $\ln c$ from $L_c$ and we 
are left with
the variable ${\tL}_c$. We note that ${\tL}_c$
is completely analogous to the
logarithmic variable $\tL$ in Eq.~(\ref{ltilde}) (e.g., we can simply set 
$c=b_0^2M^2/Q^2$). In particular, the quantitative effect of chosing
different values of the constant $c$ in ${\tL}_c$ is completely equivalent 
to the effect of using different values of the resummation scale $Q$ in    
$\tL$.
}
in Eq.~(\ref{ltilde}) has thus a twofold 
consequence \cite{Bozzi:2005wk}: 
it reduces the impact of resummation at large values of $q_T$, and it acts as
a constraint of perturbative unitarity since
it allows us to exactly recover the fixed-order value
of the total cross section 
upon integration over $q_T$ of the resummed calculation of $d\sigma/dq_T$.
(i.e., the resummed terms give a vanishing contribution to the total cross
section). 

%
%

The process dependence (as well as the factorization scale dependence) 
of the resummation factor ${\cal W}_{N}^V$ is fully encoded
in the hard function ${\cal H}_N^{V}$ on the right-hand side of 
Eq.~(\ref{wtilde}).
Since this function
does not contain 
large logarithmic terms to be resummed, it can be expanded in 
powers of $\as=\as(\mu_R^2)$ as follows:
\begin{align}
\label{hexpan}
{\cal H}_N^{V}(M,\as;M^2/\mu^2_R,M^2/\mu^2_F,M^2/Q^2)&=
\sigma_{V}^{(0)}(M)
\Bigl[ 1+ \f{\as}{\pi} \,{\cal H}_N^{V \,(1)}(M^2/\mu^2_F,M^2/Q^2) 
\Bigr. \nn \\
&+ \Bigl.
\left(\f{\as}{\pi}\right)^2 
\,{\cal H}_N^{V \,(2)}(M^2/\mu^2_R,M^2/\mu^2_F,M^2/Q^2)+\dots \Bigr] \;\;,
\end{align}
where $\sigma_{V}^{(0)}$ is the partonic cross section at the Born level.

In the case of production of the DY lepton pair $\ell \;\ell'$, 
$\sigma_{V}^{(0)}$ is the electroweak cross section of 
the process $q {\bar q} \to V \to \ell \;\ell'$, and the corresponding 
first-order coefficients ${\cal H}_{q{\bar q}\ito ab,N}^{V(1)}$
in Eq.~(\ref{hexpan})
are known \cite{Davies:1984hs}. They are:
\begin{equation}
\label{H1}
{\cal H}^{V(1)}_{q{\bar q}\ito q{\bar q},N}=
C_F\left(\f{1}{N(N+1)}-4+\f{\pi^2}{2}\right)
-\left(B^{(1)}_q+\f{1}{2} A^{(1)}_q \ln\f{M^2}{Q^2}\right)\ln\f{M^2}{Q^2}
+2\gamma^{(1)}_{qq, \, N}\, \ln\f{Q^2}{\mu_F^2}\;\; ,
\end{equation}
\begin{equation}
\label{H1off}
{\cal H}^{V(1)}_{q{\bar q}\ito gq,N}={\cal H}_{q{\bar q}\ito qg,N}^{V(1)}
=\f{1}{2(N+1)(N+2)} \, + \gamma_{qg,N}^{(1)}\, \ln\f{Q^2}{\mu_F^2}\;\; ,
\end{equation}
\begin{equation}
\label{H1gg}
{\cal H}^{V(1)}_{q{\bar q}\ito gg,N}={\cal H}^{V(1)}_{q{\bar q}\ito qq,N}=
{\cal H}^{V(1)}_{{q}{\bar q}\ito {\bar q}{\bar q},N} = 0 \;\;,
\end{equation}
where $\gamma^{(1)}_{ab,N}$ are the leading order (LO) anomalous dimensions.
The second-order coefficients
${\cal H}_{q{\bar q}\ito ab,N}^{V(2)}$ for the DY process
have not yet been computed.

We now turn to consider the finite component, 
of the cross section (see Eq.~(\ref{resplusfin})).
Since $d{\hat \sigma}^{({\rm fin.})}_{V\, ab}$ does not contain large
logarithmic terms, it 
can be computed by truncation of the perturbative series at a given fixed order
(LO, NLO and so forth).
This component is evaluated by
starting from the usual perturbative truncation of the partonic cross section
$d{\hat \sigma}_{V\, ab}$ at a given order
and subtracting the expansion of the resummed component  
$d{\hat \sigma}^{({\rm res.})}_{V\, ab}$ at
the {\em same} perturbative order (see Sect.~2.4 of Ref.~\cite{Bozzi:2005wk}).
Using this procedure, the resummed and fixed-order calculations are 
consistently
matched by avoiding double-counting of perturbative contributions
in the region of intermediate and large values of $q_T$.

The formalism that we have briefly recalled in this section defines 
a systematic `order-by-order' (in extended sense)
expansion \cite{Bozzi:2005wk} of Eq.~(\ref{resplusfin}):
it can be used to obtain predictions that contain the full information of the
perturbative calculation up to a given fixed order plus resummation of
logarithmically-enhanced contributions from higher orders.
The various
orders of this expansion are denoted\footnote{In the literature on $\qt$
resummation, other authors sometime use the same labels (NLL, NLO and so 
forth) with a meaning that is different from ours.}
as LL, NLL+LO, NNLL+NLO, etc., where the 
first label (LL, NLL, NNLL, $\dots$) refers to the logarithmic accuracy at 
small $q_T$ and the second label (LO, NLO, $\dots$) refers to the customary 
perturbative order
at large $q_T$. 
To be precise,
the NLL+LO term of Eq.~(\ref{resplusfin}) is obtained by including  
the functions $g^{(1)}$, $g^{(2)}_N$
and the coefficient ${\cal H}_N^{V(1)}$ 
in the resummed component, and by computing the finite (i.e. large-$q_T$)
component at the LO (i.e. at ${\cal O}(\as)$ for the DY process). 
At NNLL+NLO accuracy, the resummed component
includes also the function $g_N^{(3)}$ and the coefficient 
${\cal H}_N^{V(2)}$, 
while the finite component is expanded up to NLO 
(i.e. at ${\cal O}(\as^2)$ for the DY process).
It is worthwhile noticing 
that the NLL+LO (NNLL+NLO) result includes the {\em full} NLO (NNLO)
perturbative contribution in the small-$\qt$ region.
In particular,  the NLO (NNLO) result for total cross section  
is exactly recovered upon integration
over $q_T$ of the differential cross section $d \sigma/dq_T$ at NLL+LO 
(NNLL+NLO) accuracy.

In the case of the DY process, the second-order coefficient 
${\cal H}_N^{V(2)}$ is still unknown: this prevents us from performing
calculations at NNLL+NLO accuracy. 
In the following we limit ourself to presenting results of calculations at
NLL+LO accuracy\footnote{
Perturbative information with comparable accuracy at small and 
intermediate values of $q_T$ is 
implemented in 
the Monte Carlo 
event generators MC@NLO \cite{MCatNLO} and POWHEG \cite{Alioli:2008gx,Hamilton:2008pd}.}.

We note that the 
inclusion
of the function $g_N^{(3)}$ and
of the finite component at NLO  
is feasible at present. 
This procedure implements the complete NLO information at large $q_T$,
but it does not lead to a consistent systematic improvement
of the perturbative accuracy at small $q_T$. For instance, this procedure
does not recover the total cross section at NNLO: the missing NNLO (i.e.
${\cal O}(\as^2)$)
contribution to the total cross section is due to ${\cal H}_N^{V(2)}$,
and it is localized in the small-$q_T$ region.
Moreover, starting from ${\cal O}(\as^3)$, 
the contribution $\as \,g_N^{(3)}(\as L)$ in Eq.~(\ref{exponent})
and the missing contribution 
from the combined effect of $\as^2 \,{\cal H}_N^{V(2)}$ and
$L \,g^{(1)}(\as L)$ (i.e. $\as^2 \,{\cal H}_N^{V(2)} \,L \,g^{(1)}(\as L)=
\as \,{\cal H}_N^{V(2)} \as L \,g^{(1)}(\as L)$)
are of the same logarithmic order,
namely, they are both NNLL contributions ($\propto \as (\as L)^n$).

Within our formalism, the resummation factor ${\cal W}_{N}^{V}(b,M)$
is directly defined, at fixed $M$,
in the space of the conjugate variables $N$ and $b$. 
To obtain the hadronic (partonic) cross
section, as function of the kinematical variables $s$ ($\hat s$) 
and $q_T$, we have to perform inverse integral
transformations. These 
integrals
are carried out numerically.
We recall \cite{Bozzi:2005wk} that the resummed form factor 
$\exp \{{\cal G}_N(\as(\mu_R^2),{\widetilde L})\}$
(more precisely, each of the
functions $g_N^{(i)}(\as {\widetilde L})$ in Eq.~(\ref{exponent})) 
is singular at 
the values of $b$ where $\as(\mu_R^2) {\widetilde L} \geq \pi/\beta_0$ 
($\beta_0$ is the first-order coefficient of the QCD $\beta$ function). 
When performing the inverse Fourier (Bessel)
transformation with respect to the impact parameter $b$ 
(see Eq.~(\ref{resum})), we deal with this
singularity by using the regularization prescription of
Refs.~\cite{Laenen:2000de,Kulesza:2002rh}:
the singularity is avoided by deforming the 
integration contour in the complex $b$ space.

The singularity of the resummed form factor occurs at large values of the impact
parameter: $b \gtap 1/\Lambda_{QCD}$, where $\Lambda_{QCD}$ is 
the momentum scale of the Landau pole of the perturbative running coupling
$\as(q^2)/\pi \sim [\beta_0 \ln(q^2/\Lambda_{QCD}^2)]^{-1}$.
This singularity signals the onset of non-perturbative phenomena at very large
values of $b$ or, equivalently, in the region of very small transverse momenta.
The regularization prescription that we use has to be regarded as  
a `minimal prescription' \cite{Catani:1996yz,Laenen:2000de} 
within a purely perturbative framework. The prescription leaves unchanged the
perturbative result to any (and arbitrarily-high) fixed order in $\as$, it 
does not require any infrared  cut-off, and it can be implemented without
introducing an explicit model of non-perturbative (NP) contributions.
Owing to these features, the prescription is suitable to examine the 
perturbative effects and the ensuing perturbative uncertainty.
This does not imply that NP contributions are small and can be neglected.
We comment on NP effects at the end of Sect.~\ref{sec:resu}.

\section{$Z$ production at Tevatron energies: numerical results}
\label{sec:results}

In this section we present a selection of our numerical results, by considering
the production of Drell--Yan $e^+e^-$ pairs in $p{\bar p}$ collisions at 
Tevatron energies. The numerical results are also compared
with the data collected by the CDF and D0 experiments 
at the energies
$\sqrt{s}$ = 1.8~TeV (Run I) \cite{Affolder:1999jh,Abbott:1999yd} and 
$\sqrt{s}$ = 1.96~TeV (Run II, D0 only) \cite{:2007nt}.

As for the electroweak couplings, we use the scheme where the input parameters
are $G_F$, $m_Z$, $m_W$ and $\alpha(m_Z)$. In particular, we use the values 
$G_F = 1.16639\times 10^{-5}$~GeV$^{-2}$, $m_Z = 91.188$~GeV, $m_W = 80.419$~GeV and
$\alpha(m_Z) = 1/128.89$.
Our calculation implements 
the decays $\gamma^* \to e^+e^-$ and $Z^* \to e^+e^-$
at fixed value of the invariant mass of the $e^+e^-$ pair. In particular, we
include the effects of the $\gamma^*\,Z$ interference and of the 
finite width of the $Z$ boson.
Nonetheless, 
the numerical results of this section are obtained by simply using the 
narrow-width approximation 
and neglecting the photon contribution. We find that this approximation works 
to better than 1\% accuracy in the inclusive regions of $e^+e^-$ invariant
mass that are covered by the CDF and D0 data.
We recall that the measured $q_T$ spectra are inclusive over 
the following ranges of $e^+e^-$ invariant mass: 
66--116~GeV \cite{Affolder:1999jh}, 75--105~GeV \cite{Abbott:1999yd} and
70--110~GeV \cite{:2007nt} .

\subsection{Fixed-order results}
\label{sec:fo}

We start the presentation of the numerical results by 
considering QCD calculations at fixed order.
To compute the LO and NLO hadronic cross section we use
the MRST2002 LO \cite{Martin:2002aw} 
and MRST2004 NLO \cite{Martin:2004ir}
parton distribution functions,
with $\alpha_S$ evaluated at 1 and 2 loops,
respectively. As for renormalization and factorization scale, we choose $\mu_F=\mu_R=m_Z$ as central value.
The fixed-order predictions for the $q_T$-spectrum are obtained 
by using a numerical program that implements the analytical results of 
Refs.~\cite{Ellis:1981hk,Arnold:1988dp,Gonsalves:1989ar}. Similar numbers
are obtained by using the Monte Carlo code of the MCFM package \cite{mcfm}.

\begin{figure}[htb]
\begin{center}
\includegraphics[width=.8\textwidth]{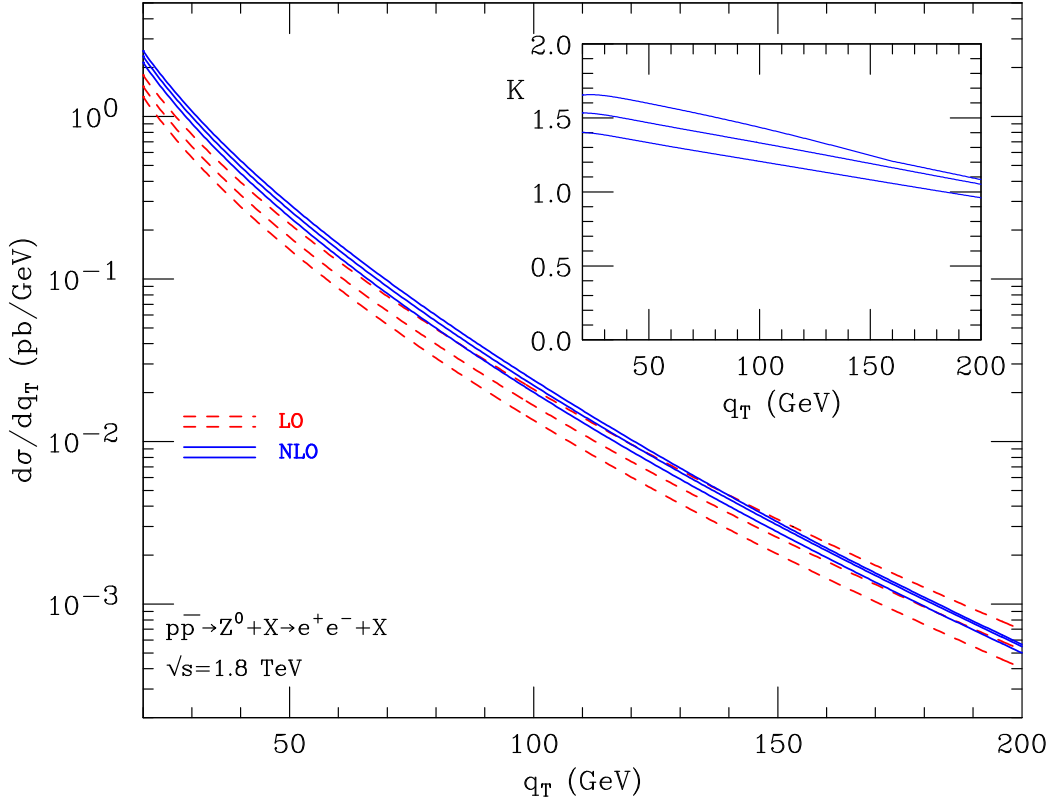}
\caption{
\it The $q_T$ spectrum of the Drell-Yan $e^+ e^-$ pairs produced in 
$p\bar{p}$ collisions at the Tevatron Run I: 
theoretical predictions at LO (dashed lines) and NLO (solid lines).
The inset plot shows the ratio K of the NLO and LO results.}
\label{fig:fo}
\end{center}
\end{figure}

In Fig.~\ref{fig:fo} we plot the $q_T$ spectrum of Drell--Yan $e^+ e^-$ pairs
at the Tevatron Run I ($\sqrt{s}=1.8$~GeV) at LO and NLO accuracy.
The bands are obtained by varying independently the
factorization ($\mu_F$) and renormalization ($\mu_R$) scales 
in the range $0.5 m_Z\leq \mu_F,\mu_R\leq 2 m_Z$, with the constraint
$0.5 \leq \mu_F/\mu_R \leq 2$. 
At the LO level, the scale dependence is about $\pm$25\%  at large $q_T$,
it decreases as $q_T$ decreases, and it becomes about
$\pm$15\% at $q_T\sim 20$~GeV. 
The scale dependence at NLO is about $\pm$8\% at $q_T\sim 20$~GeV, with 
a slight reduction at large $q_T$.
We note 
that the LO and NLO bands do not overlap in the region where $q_T\ltap 70$~GeV.
This proves that, in this region of transverse momenta,
the size of the band obtained through scale variations at LO
definitely
underestimates the theoretical uncertainty due to the missing NLO corrections.

The inset plot of Fig.~\ref{fig:fo} shows the $K$-factor, obtained by normalizing the NLO band 
with the LO result at $\mu_F=\mu_R=m_Z$.
The impact of the NLO corrections, at central values of the scales, ranges
from about +10\% at $q_T\sim 200$~GeV to about +50\% at $q_T\sim 20$~GeV.

\begin{figure}[htb]
\begin{center}
\includegraphics[width=.8\textwidth]{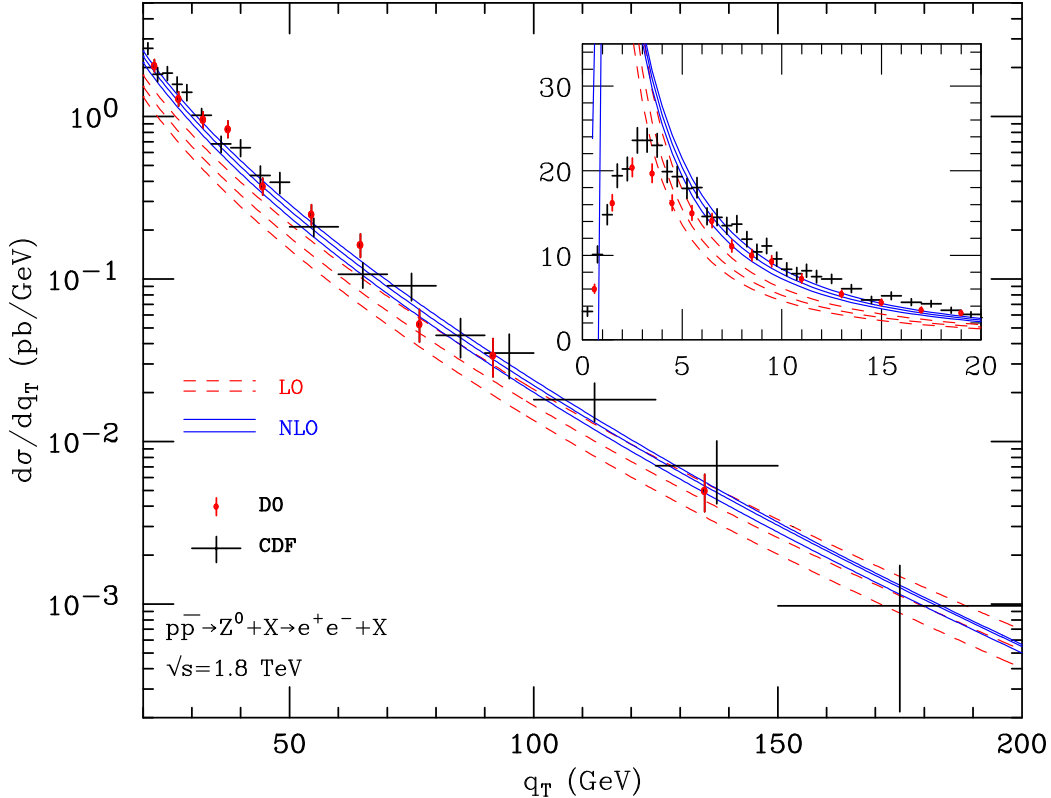}
\caption{
\it The $q_T$-spectrum of the Drell-Yan $e^+ e^-$ pairs produced in 
$p\bar{p}$ collisions at the Tevatron Run I.
The data are from Refs.~\cite{Affolder:1999jh,Abbott:1999yd}.
Theoretical results are shown at LO (red dashed lines) and NLO 
(blue solid lines) including scale variations. 
}
\label{fig:forun1}
\end{center}
\end{figure}

In Fig.~\ref{fig:forun1} we compare the results of the fixed-order calculations
with CDF and D0 data.
The experimental error bars reported in Fig.~\ref{fig:forun1} include 
statistical and systematic contributions, but they do not include the overall
normalization uncertainty due to the luminosity measurement.
The CDF and D0 luminosity uncertainties are $\pm$3.9\% and $\pm$4.4\%,
respectively.
The corresponding cross sections
are $\sigma_{\rm CDF}=248\pm 11$~pb \cite{Affolder:1999jh} and 
$\sigma_{\rm D0}=221\pm 11$~pb \cite{Abbott:1999tt}, where the errors are
dominated by the luminosity uncertainties.
We note that, using the MRST2004 parton distribution functions 
\cite{Martin:2004ir} and including the effect of scale variations, 
the values of the QCD cross section at NLO and NNLO are
$\sigma_{NLO}=226 \pm 5$~pb and $\sigma_{NNLO}=236 \pm 2$~pb, respectively.

From Fig.~\ref{fig:forun1}
we see that the NLO result is in 
agreement with the experimental data over
a wide region of transverse momenta, from large to relatively-small values
of $q_T$.
In particular, in the region 20~GeV$\ltap q_T \ltap 70$~GeV, the sizeable
increase of the LO result produced by the NLO corrections is important
to achieve the agreement between the data and the NLO calculation.
In the small $q_T$ region, 
which is shown in the inset plot,
the LO and NLO calculations do not describe the data. This is not unexpected
since, when $q_T \to 0$, the LO and NLO cross section eventually 
diverges to $+\infty$ and $-\infty$, respectively.
This is the region where the effects 
of soft-gluon resummation are essential and have to be taken into account.

\begin{figure}[htb]
\begin{center}
\includegraphics[width=.9\textwidth]{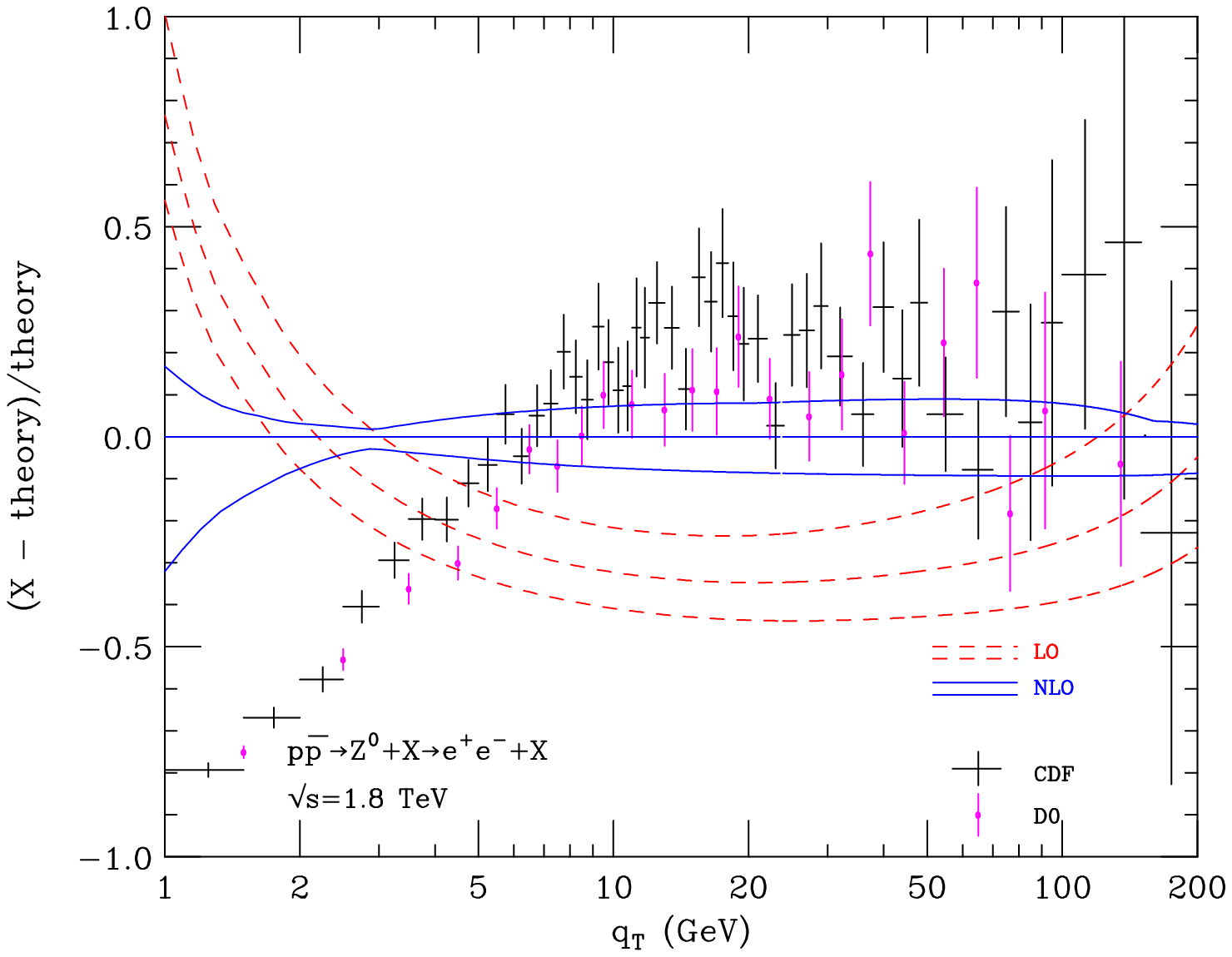}
\caption{
\it Fractional difference of fixed-order predictions and Tevatron Run I data with respect to the NLO result at $\mu_F=\mu_R=m_Z$ (see Eq. (\ref{fracnlo})).
}
\label{fig:rforun1}
\end{center}
\end{figure}

A more effective comparison between the fixed-order calculations and the data
can be performed by considering the fractional difference 
$(X - {\rm theory})/{\rm theory}$. We choose the NLO result at central value  
of the scales as `reference theory' and we define the following
fractional difference (see Fig.~\ref{fig:rforun1}):
\begin{equation}
\label{fracnlo}
\frac{(d\sigma/dq_T)_{X} - (d\sigma/dq_T)_{NLO}(\mu_F=\mu_R=m_Z)}
{(d\sigma/dq_T)_{NLO}(\mu_F=\mu_R=m_Z)} \; \; ,
\end{equation}
where the label $X$ refers to either the LO and NLO results, including scale 
variations (dashed and solid curves in Fig.~\ref{fig:rforun1}), 
or the experimental data.

The perturbative QCD predictions at NLO have an associated theoretical
uncertainty due to missing higher-order terms. By comparing the numerical
results at LO and NLO, we can try to consistently estimate this uncertainty.
At large values of $q_T$, the LO and NLO bands overlap. In this region
of transverse momenta, we can thus use the scale variation band as uncertainty
estimate: we obtain that the NLO predictions have a perturbative uncertainty of
about $\pm$8\%. As $q_T$ decreases below the value $q_T \sim 70$~GeV, 
the LO and NLO bands do not overlap (see Fig.~\ref{fig:rforun1}), 
and this signals that scale variations
(up to NLO) tend to underestimate the effect of higher-order contributions.
To obtain a more reliable estimate of the uncertainty of the NLO central value,
we assign it a theoretical error as given by its difference with respect
to its closest value in the LO band (i.e., the value on the upper curve 
of the LO band in Fig.~\ref{fig:rforun1}). Using this procedure, 
we obtain a NLO uncertainty 
that increases as $q_T$ decreases, and that reaches the value 
of about $\pm$20\% at $q_T \sim 20$~GeV. In the region of smaller values of
$q_T$, 
the LO and NLO results show a pathological behaviour.
This behaviour, which is discussed below,
prevents us from making a sensible quantitative estimate of the
theoretical uncertainty of the NLO predictions. We can only draw a qualitative
conclusions: the uncertainty of the NLO predictions systematically increases as 
$q_T$ decreases.

We know that, in the small-$q_T$ region, the convergence of the fixed-order
perturbative expansion is spoiled by the presence of large logarithmic
corrections. This behaviour is clearly seen in Fig.~\ref{fig:rforun1}
by comparing the LO and NLO results at $q_T \sim 1$~GeV. 
We also recall that, as $q_T \to 0$,
the LO cross section diverges to $+\infty$ whereas the NLO cross section 
diverges to $-\infty$. Since the NLO corrections increase the LO results at 
large $q_T$, the LO and NLO cross sections have to coincide at some
intermediate value of $q_T$, before reaching the small-$q_T$ region where 
their different divergent behaviour sets in.  
The numerical 
agreement of the LO and NLO results occurs in the region 
where $q_T \sim$ 2--3~GeV
(Fig.~\ref{fig:rforun1}). 
Since this region is so close to that where
the convergence of the fixed-order is definitely spoiled,
it cannot be regarded as a region where the order-by-order expansion is 
well-behaved.
Therefore,
the systematic decrease of the difference between the LO and NLO cross 
sections when $q_T$ varies from about 20~GeV to 2--3~GeV 
is simply driven by the sickness of the LO and NLO results at
smaller $q_T$.
This decrease cannot be interpreted
as an increase of the theoretical accuracy of the NLO predictions.
The behaviour of the LO and NLO results below $q_T \sim 20$~GeV signals  
the necessity to include the effect of higher-order contributions
and, eventually, of resummed calculations.

Having estimated the perturbative uncertainty of the NLO predictions, we can 
add some comments on the comparison with the experimental data 
(see Fig.~\ref{fig:rforun1}).
We consider the region where  $q_T\gtap 20$~GeV, since at smaller values of 
$q_T$ the NLO calculation looses predictivity. Throughout this region,
the data agree with the NLO predictions.
The experimental errors are typically larger (smaller) 
than the NLO uncertainty when $q_T\gtap 70$~GeV 
($20~{\rm GeV} \ltap q_T \ltap 30$~GeV).
The experimental errors and the corresponding NLO errors
overlap, with the sole exception of a couple of D0 data points. 
We note that part of the differences between data and theory have a systematic
component due to the luminosity uncertainties ($\pm$3.9\% at CDF, and 
$\pm$4.4\% at D0)
and to the values of the total cross
sections. The NLO predictions for $d\sigma/dq_T$ correspond to the
value\footnote{This value is obtained by convoluting the NNLO partonic cross
sections with NLO parton distributions (see the comment above
Eq.~(\ref{norsigma})).}
$\sigma=233$~pb of the total cross section; this value is about 6\%
smaller than the CDF value, and it is about 5\% larger than the D0 value.
Therefore, by inspection of Fig.~\ref{fig:rforun1}, we see that considering 
the normalized distribution $\frac{1}{\sigma} \frac{d\sigma}{dq_T}$
(i.e. the shape of the $q_T$ spectrum)
would lead to an overall improvement of the agreement between the CDF data, the
D0 data and the NLO predictions.

The D0 collaboration has performed a measurement \cite{:2007nt}
of the normalized 
$q_T$ distribution, $\frac{1}{\sigma} \frac{d\sigma}{dq_T}$, from data at the
Tevatron Run II (${\sqrt s}=1.96$~GeV). To obtain fixed-order QCD predictions
for the normalized distribution, we have to consistently compute 
$d\sigma/dq_T$ and ${\sigma}$. The cross section $d\sigma/dq_T$ at NLO (LO)
is computed by convoluting the corresponding NLO (LO) partonic cross sections
with NLO (LO) parton distributions. 
These differential partonic cross
sections at NLO (LO) contribute to the total partonic cross sections
at NNLO (NLO).
Therefore, to normalize $d\sigma/dq_T$ at NLO (LO), 
the total cross section $\sigma$ is computed by convoluting 
the total partonic cross sections at NNLO (NLO) with NLO (LO) parton
distributions. In conclusion, our fixed-order calculations 
of the normalized $q_T$ distribution are obtained by using the following
relation:
\begin{equation}
\label{norsigma}
\left(\f{1}{\sigma}\f{d\sigma}{dq_T}\right)_{(N)LO}(\mu_F,\mu_R)\equiv
\f{1}{\sigma_{(N)NLO}(\mu_F,\mu_R)} 
\left(\f{d\sigma}{dq_T}
\right)_{\!(N)LO}\!\!\!\!\!\!\!\!\!\!\!\!\!(\mu_F,\mu_R)\;\;,
\end{equation}
where the two factors, $1/\sigma$ and $d\sigma/dq_T$, on the right-hand side
are evaluated with the same parton distributions and at the same values of the
renormalization and factorization scales.

\begin{figure}[htb]
\begin{center}
\includegraphics[width=.8\textwidth]{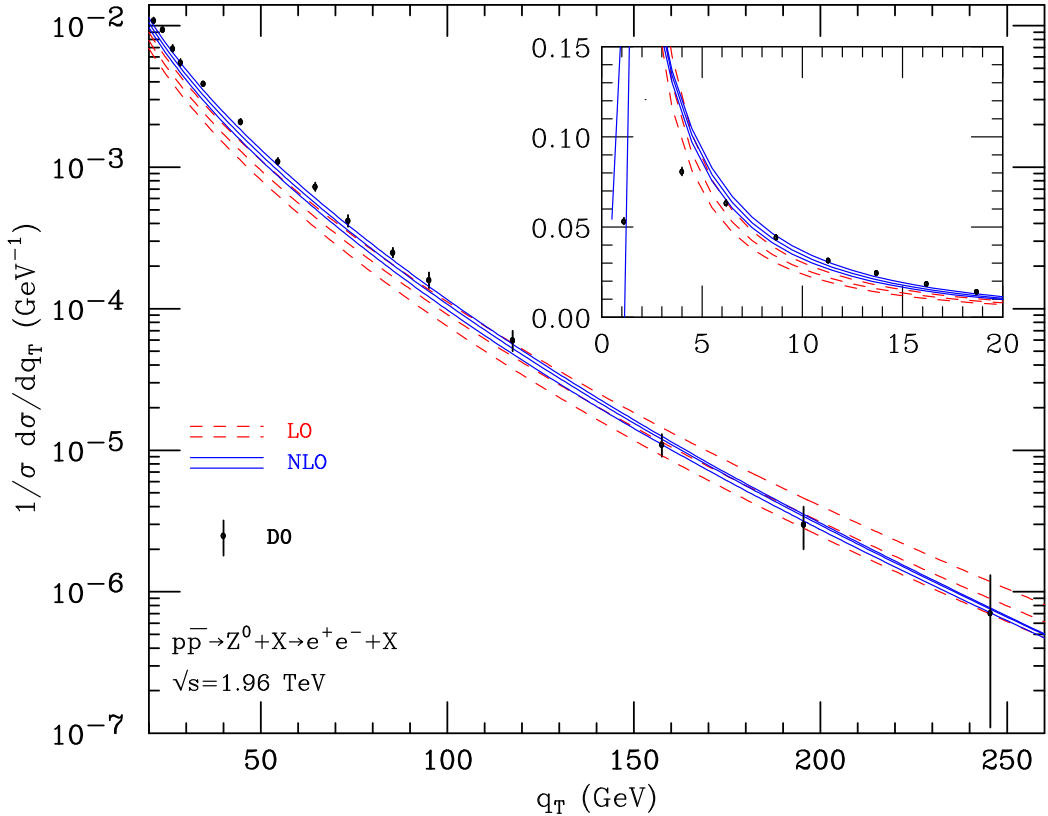}
\caption{
\it The $q_T$-spectrum of the Drell-Yan $e^+ e^-$ pairs produced in 
$p\bar{p}$ collisions at the Tevatron Run II,
normalized to the total cross section.
The data are from Ref.~\cite{:2007nt}.
Theoretical results are shown at LO (red dashed lines) and NLO 
(blue solid lines) including scale variations. 
}
\label{fig:forun2}
\end{center}
\end{figure}

\begin{figure}[htb]
\begin{center}
\includegraphics[width=.9\textwidth]{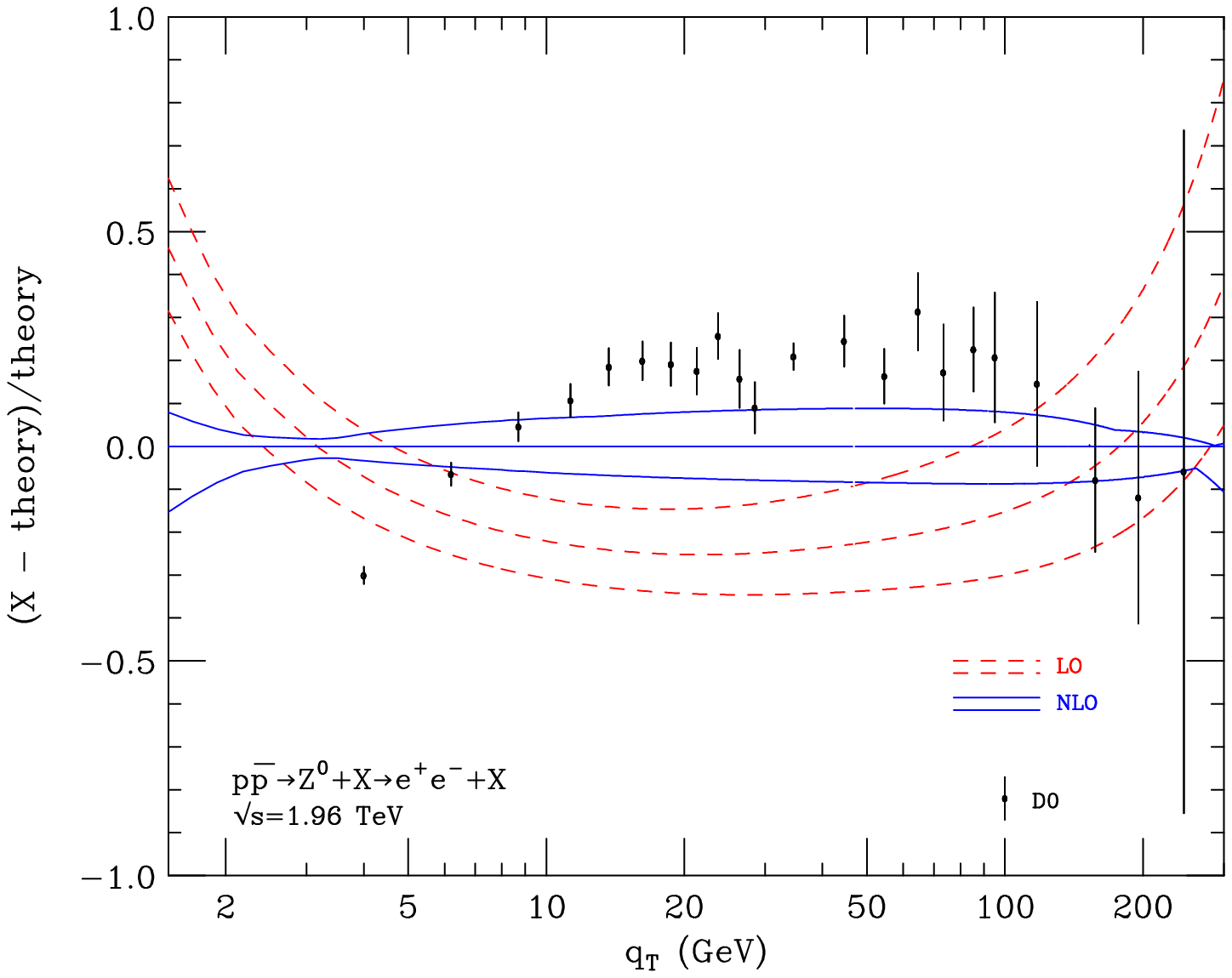}
\caption{
\it Same as Fig.~\ref{fig:rforun1} for $1/\sigma\, (d\sigma/ dq_T)$ at the Tevatron Run II.
}
\label{fig:rforun2}
\end{center}
\end{figure}

The D0 data and the LO and NLO results for the normalized distribution 
$\frac{1}{\sigma} \frac{d\sigma}{dq_T}$ at the
Tevatron Run II are presented in Fig.~\ref{fig:forun2}. The LO and NLO
bands are obtained by varying $\mu_F$ and $\mu_R$ in the same range 
as in the fixed-order calculations at Run I.
The fractional difference $(X - {\rm theory})/{\rm theory}$ at Run II
is shown in Fig.~\ref{fig:rforun2}; this fractional difference differs from 
that in Eq.~(\ref{fracnlo}) by the sole replacement of
$\frac{d\sigma}{dq_T}$ with $\frac{1}{\sigma} \frac{d\sigma}{dq_T}$.

Comparing Figs.~\ref{fig:forun1} and \ref{fig:rforun1} with 
Figs.~\ref{fig:forun2} and \ref{fig:rforun2}, we can see that the overall 
features
of the fixed-order results are unchanged in going from Run I to Run II.
The main quantitative differences are due to the fact that, at Run II, we are
considering the normalized $q_T$ distribution. 
The NLO corrections to $\frac{1}{\sigma} \frac{d\sigma}{dq_T}$ are smaller than
the NLO corrections to $\frac{d\sigma}{dq_T}$. The scale dependence of the
fixed-order results is only marginally reduced by considering 
$\frac{1}{\sigma} \frac{d\sigma}{dq_T}$ rather than $\frac{d\sigma}{dq_T}$.

The perturbative uncertainty of the NLO predictions at Run II
can be estimated in the same way as at Run I. We conclude that the NLO error
increases from 
about $\pm 6$--8\% in the region where $q_T \gtap 50$~GeV
(see the size of the NLO band in Fig.~\ref{fig:rforun2})
to about $\pm 15$\% at $q_T \sim 20$~GeV (see the upper value of the 
LO band in Fig.~\ref{fig:rforun2}). At smaller values of $q_T$ the NLO
calculation looses predictivity.
In the region where $q_T \gtap 90$~GeV, the D0 data agree with the NLO
predictions. In the region where $20~{\rm GeV} \ltap q_T \ltap 90$~GeV,
the experimental errors are typically smaller than the NLO uncertainty;
in this region, three data points overshoot the NLO predictions 
(they differ by about two standard deviations from the upper value of the 
NLO uncertainty band).

\setcounter{footnote}{1}

We add a comment on the fixed-order calculations presented in the papers
of the D0 collaboration. The labels ``Fixed-order $({\cal O}(\as^2))$''
in Ref.~\cite{Abbott:1999yd}
and ``NNLO'' in Ref.~\cite{:2007nt} refer to perturbative calculations that use
the same NLO partonic cross sections of our NLO calculations. The differences
between these calculations can be due to the use of different parton
distributions and of different renormalization and factorization scales.
In the region where $q_T \gtap 20$~GeV, the fixed-order results
of Refs.~\cite{Abbott:1999yd,:2007nt} are systematically smaller than
our NLO central values:
the differences can reach the level of about
12\% in the case of the ``Fixed-order $({\cal O}(\as^2))$'' result
(compare Fig.~26 of the second paper in Ref.~\cite{Abbott:1999yd}
with our Fig.~\ref{fig:rforun1}) and of about 7\%
in the case of the ``NNLO'' result (compare Fig.~2b in Ref.~\cite{:2007nt} 
with our Fig.~\ref{fig:rforun2}). 
We note that the differences tend 
to reduce the agreement with the D0 data;
we also note that these differences 
are consistent with our estimate of the perturbative uncertainty of 
the NLO predictions. This observation underlines the relevance of quantifying
the uncertainty of the perturbative QCD predictions.

\subsection{Resummed results}
\label{sec:resu}

In the following we present our resummed results at NLL+LO accuracy 
and we compare them with the Tevatron data. 
To compute the hadronic cross sections,
we use the MRST2004 NLO parton distributions \cite{Martin:2004ir}, 
with $\as$ evaluated at 2-loop order.

In the small-$q_T$ region, the use of NLO parton distributions 
is fully consistent with the NLL+LO accuracy of the partonic cross section.
Indeed, at small values of $q_T$, the NLL+LO partonic cross section includes
the complete perturbative expansion up to NLO (i.e., ${\cal O}(\as)$) and the
resummation of the LL and NLL terms. The use of NLO parton distributions
is justified also at intermediate values of $q_T$, where the calculation of the
partonic cross section is driven by the small-$q_T$ resummation and constrained
by the value of the total cross section at NLO.

The resummed calculation depends on the factorization and 
renormalization scales and on the resummation scale $Q$. 
As in fixed-order calculations, 
we choose $\mu_F=\mu_R=m_Z$ as central value. Factorization 
and renormalization scale uncertainties are computed as described in 
Sect.~\ref{sec:fo},
by considering  
variations of $\mu_F$ and $\mu_R$ by a factor of two around (above and below)
the central value. A similar procedure is applied to the resummation scale:
we choose $Q=m_Z/2$ as central value and consider scale variations in the 
range $m_Z/4 < Q < m_Z$.
As discussed below, we regard this central value of the resummation scale
and the corresponding range $m_Z/4 < Q < m_Z$ as sufficiently conservative
(e.g., more
conservative than the range $m_Z/2 < Q < 2m_Z$) from a theoretical 
viewpoint\footnote{Also in the case of Higgs boson
production, we used \cite{Bozzi:2005wk} the range $m_H/4 < Q < m_H$ 
($m_H$ being the mass of the Higgs boson) to estimate the resummation scale
uncertainty.}.

The resummed logarithmic terms depend on $Q$ through 
the variable $\tL$ in Eq.~(\ref{ltilde}). This implies that the logarithmic
contributions are mostly effective in the $b$-space region where $bQ \gtap 1$,
which corresponds to the transverse-momentum region where $q_T \ltap Q$.
By decreasing $Q$, the resummation effects are depleted in the region where
$q_T \gtap Q$ and enhanced in the region where $q_T \ltap Q$.
The bulk of the $Z$ production cross section and, thus,
the main effect of the logarithmic terms are located at values of 
$q_T$  that are certainly smaller than $m_Z$: indeed, we observe that
the average transverse 
momentum of the $Z$ boson is of the order of $\as(m_Z) \,m_Z$.
Therefore, it is physically sensible to use a central value of $Q$ that is 
smaller than $m_Z$. Nonetheless, too small values of $Q$ have to be avoided.
As we have pointed out in Sect.~\ref{sec:fo}, the fixed-order perturbative
expansion shows instabilities (due to higher-order logarithmic corrections)
in the region where $q_T \ltap 20$~GeV. Therefore, in our NLL+LO calculation
we should exclude values of $Q$ that are smaller than about 20~GeV.
In this respect, a value of $Q$ as low as $Q \sim m_Z/4$ can be
regarded as a conservative value from a perturbative viewpoint. 
The NLL+LO calculation with such a value of $Q$ will be closer to the 
corresponding fixed-order calculation throughout region of intermediate values 
of $q_T$ where the fixed-order expansion is relatively well behaved.


\begin{figure}[htb]
\begin{center}
\includegraphics[width=0.9\textwidth]{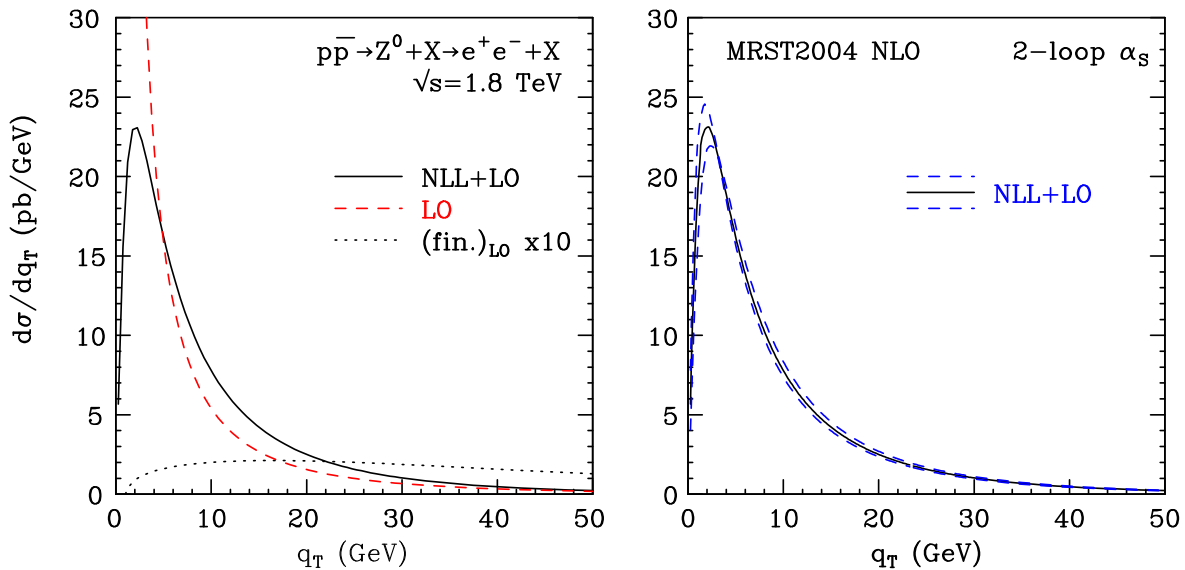}
\caption{
\it The NLL+LO $q_T$ spectrum at the Tevatron Run I.
}
\label{fig:nllnlo}
\end{center}
\end{figure}

The NLL+LO $q_T$ spectrum
at the Tevatron ($\sqrt{s}=1.8$~TeV) is presented in Fig.~\ref{fig:nllnlo}.
In the left panel, the NLL+LO 
result (solid line) at
the default scales ($\mu_F=\mu_R=m_Z$, $Q=m_Z/2$) is compared with the
corresponding
\footnote{Here and in the inset plot of Figs.~\ref{fig:nllnloadep}, 
the LO result refers to the
convolution of the partonic cross section at LO with the parton distributions at
NLO. This LO result thus differs from the customary LO calculation, which uses
parton distributions at LO and is presented in
Sect.~\ref{sec:fo}. Incidentally, we note that the difference produced by using
NLO vs. LO parton distributions is much smaller than the scale uncertainty of 
the corresponding results. We find that the difference is below the level of
about $\pm 2$\% ($-6$\%) if  $20~\rm{GeV} \ltap q_T \ltap 140~\rm{GeV}$
($140~\rm{GeV} \ltap q_T \ltap 200~\rm{GeV}$).}
LO result (dashed line).
We observe that soft-gluon 
resummation leads to a well-behaved distribution: it vanishes as $q_T\to 0$, has
a kinematical peak at $q_T\sim 2$~GeV, and tends to the corresponding LO result
at larger values of $q_T$.
The LO finite component of the spectrum (see Eq.~(\ref{resplusfin})),
rescaled by a factor of 10 to make it more visible, is also shown 
for comparison (dotted line).
This component
smoothly vanishes
as $q_T\to 0$ and 
gives a small contribution to the NLL+LO result in the low-$q_T$ region.
The contribution is smaller than 1\% at the peak
and becomes more important as $q_T$ increases: it is about 8\% at $q_T\sim
20$~GeV, about 20\% at $q_T\sim 30$~GeV and about 60\% at $q_T\sim 50$~GeV.
A similar quantitative behaviour is observed by considering the contribution of
the LO finite component to the LO result;
the contribution is about 13\% at $q_T\sim 20$~GeV, 
about 30\% at $q_T\sim 30$~GeV and about 75\% at $q_T\sim 50$~GeV.
In the region of intermediate values of $q_T$ (say, around 20~GeV),
the difference between the NLL+LO and LO results is much larger than the size
of the LO finite component. This difference is produced by the logarithmic
terms (at NLO and beyond NLO) that are included in the resummed calculation at
NLL accuracy. At large values of $q_T$ the contribution of
the LO finite component sizeably increases. This behaviour indicates that  
the logarithmic terms are no longer dominant and that the resummed
calculation cannot improve upon the predictivity of the fixed-order expansion.

In the right panel of Fig.~\ref{fig:nllnlo} we show the scale dependence 
of the NLL+LO result.
The band (dashed lines) is obtained by varying the renormalization and 
factorization scales as described in Sect.~\ref{sec:fo}. 
Although $\mu_R$ and $\mu_F$ are varied independently, we find that the 
dependence on $\mu_R$ dominates at any values of $q_T$.
In the region of small and intermediate values of $q_T$, the scale dependence of
the NLL+LO result is definitely much smaller than the difference between the
NLL+LO and LO results.
The scale dependence of the resummed result is about $\pm 5$\%
at the peak, and it increases to $\pm 9$\% at $q_T\sim 50$~GeV.

\begin{figure}[htb]
\begin{center}
\includegraphics[width=.8\textwidth]{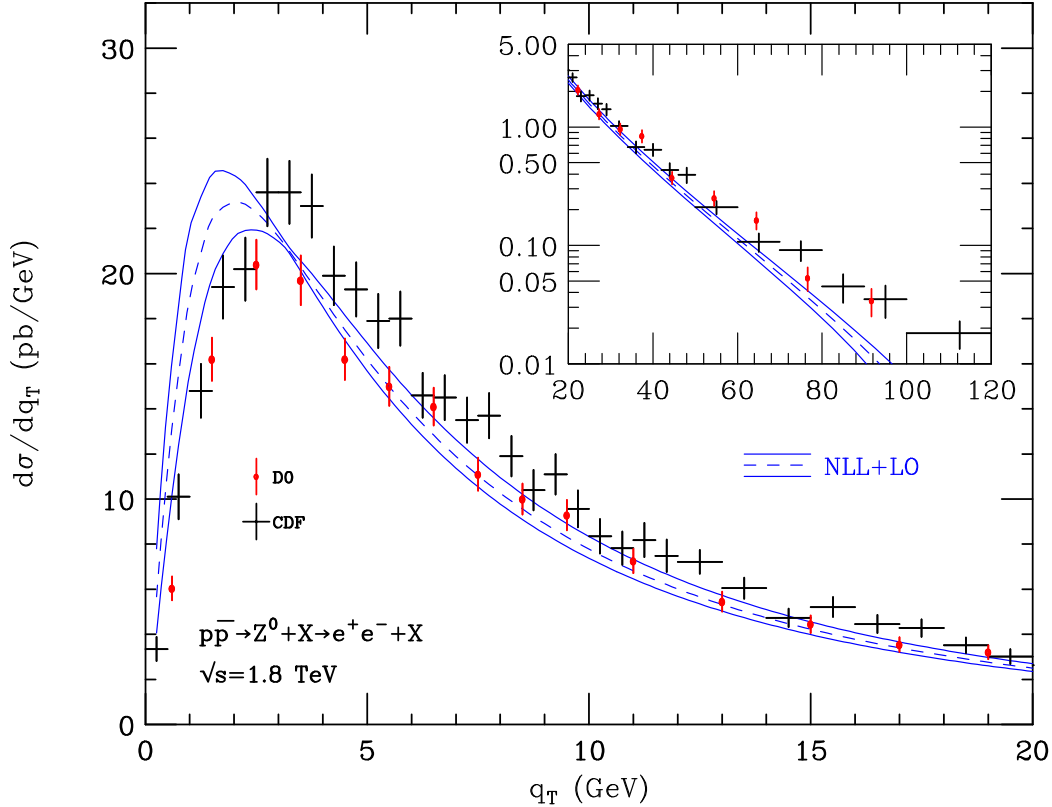}
\caption{
\it Comparison of Tevatron Run I data with the NLL+LO result including
variations of the scales $\mu_F$ and $\mu_R$.
}
\label{fig:nllnlorun1}
\end{center}
\end{figure}

\begin{figure}[htb]
\begin{center}
\includegraphics[width=.8\textwidth]{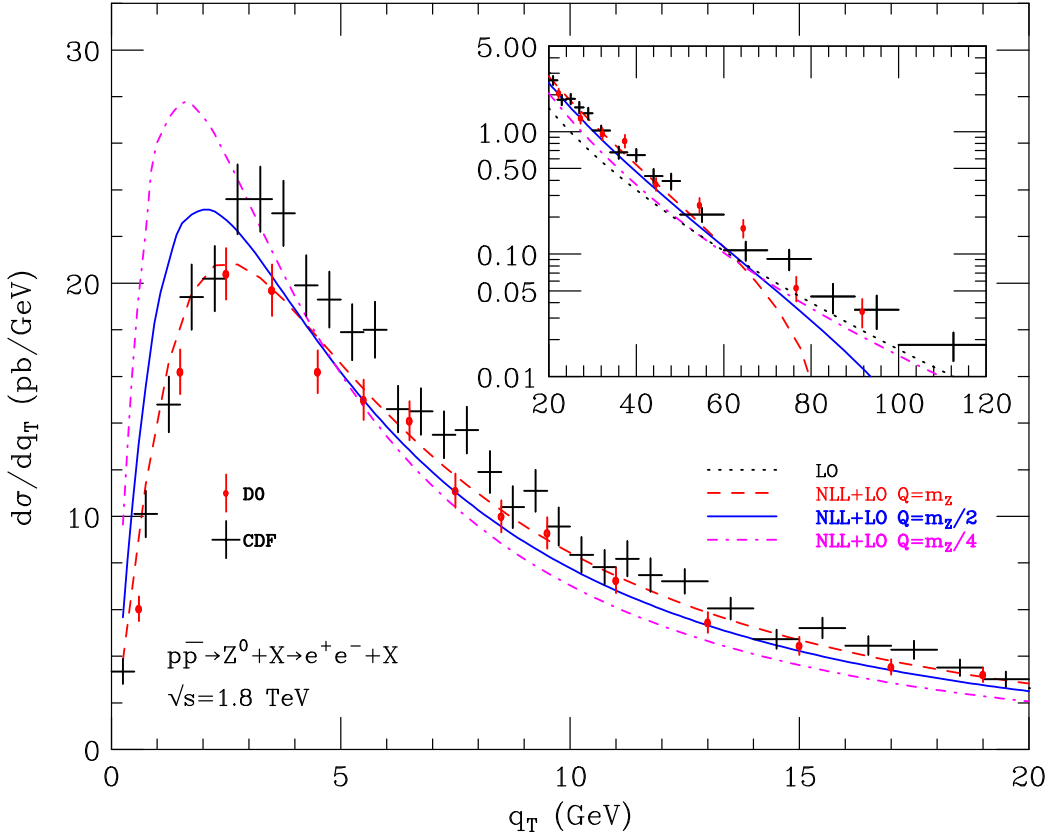}
\caption{
\it NLL+LO results at the Tevatron Run I for different values of the resummation scale $Q$.
}
\label{fig:nllnloadep}
\end{center}
\end{figure}

The renormalization/factorization scale dependence of the NLL+LO spectrum
(the band enclosed by the solid lines)
is more clearly visible in Fig.~\ref{fig:nllnlorun1}. Here
the CDF and D0 data at Tevatron Run~I are also superimposed on the NLL+LO
result. There is an overall agreement between the data and the resummed
calculation in the region from small to intermediate values of $q_T$.
In particular, the agreement tends to improve by decreasing the value of
$\mu_R$ from $\mu_R=m_Z$ to $\mu_R=m_Z/2$. The inset plot shows the region
of intermediate and large values of $q_T$. At large $q_T$, the NLL+LO
result deviates from the data, and the deviation increases as $q_T$ increases.
As previously observed on purely theoretical grounds, the NLL+LO calculation
looses predictivity in the large-$q_T$ region. The lost of predictivity is 
also signalled by the systematic increase of the scale dependence, whose size 
is about $\pm 9$\% at $q_T\sim 50$~GeV and becomes about 
$\pm 22$\% at $q_T\sim 90$~GeV. As we shall see in the next figure, at large
values of $q_T$, the dependence on the resummation scale $Q$ is even stronger
than the dependence on $\mu_R$ and $\mu_F$.

In Fig.~\ref{fig:nllnloadep} we show how the NLL+LO result depends on the
resummation scale $Q$. We fix $\mu_R=\mu_F=m_Z$, and we present the result
of the NLL+LO calculation at three different values of the resummation scale: 
$Q=m_Z$ (dashed line), $Q=m_Z/2$ (solid line) and $Q=m_Z/4$ (dot-dashed line).
In the region of small and intermediate values of $q_T$, large values of $Q$
($Q\sim m_Z, m_Z/2$) lead to a better agreement with the experimental data. 
In this region, the $q_T$ spectrum at NLL+LO becomes softer by decreasing the
value of $Q$.
This behaviour is not unexpected.
The resummed calculation cures the instabilities of the fixed-order
calculations by implementing the physical transverse-momentum smearing 
produced by soft multiparton radiation. The size of the $q_T$ region
where the smearing takes place is controlled by the value of $Q$. 
By increasing $Q$, the resummation smearing is extended to larger values of 
$q_T$, and the $q_T$ spectrum becomes harder.

The inset plot of Fig.~\ref{fig:nllnloadep} refers to the
region of intermediate and large values of $q_T$. We 
present the NLL+LO results and
the corresponding LO result (dotted line), which does not depend on $Q$.
Considering large values of $Q$ ($Q\sim m_Z, m_Z/2$), the NLL+LO results
follow the data up to $q_T\sim 50$~GeV and deviates from the data at larger 
values of $q_T$. At large $q_T$, the deviation decreases if we consider  
the NLL+LO calculation with smaller values of $Q$. 
We also notice that the LO result is approximated better by 
the NLL+LO calculation if $Q$ is smaller. This fact is not surprising.
Varying $Q$, we smoothly set the transverse-momentum
scale below which the resummed logarithmic terms are mostly effective; 
if $Q$ is smaller,
the resummation effects are confined to a range of smaller values of $q_T$.
Independently of these
observations, the sizeable 
$Q$ dependence of the resummed
calculation at large $q_T$ confirms the lost of predictivity of the 
NLL+LO result in this transverse-momentum region.

The variations of the NLL+LO cross section produced by varying the
resummation scale give further insight on the size of yet uncalculated 
higher-order logarithmic contributions at small and intermediate values of 
$q_T$.
To quantify the resummation scale uncertainty on
the cross section,
we choose $Q=m_Z/2$ as central value and vary $Q$ between $m_Z$ and $m_Z/4$. 
We find that the uncertainty is about $\pm 12$\% in the region of the peak,
it decreases in the region around $q_T\sim 5$~GeV, and then it increases 
to about $\pm 15$\% in the region where $q_T\sim 20$~GeV.

The integral over $q_T$ of the NLL+LO spectrum is in agreement (for any values 
of $\mu_R, \mu_F$ and $Q$) with the value of
the NLO total cross section to better than 1\%,
thus checking
the numerical accuracy of our code. 
We also note that the large-$q_T$ region gives a little contribution to
the total cross section; therefore, 
the total cross section
constraint mainly acts as a perturbative constraint on the NLL+LO spectrum 
in the region from intermediate
to small values of $q_T$. To confirm this statement at the quantitative level,
we consider the cross section ratio $\sigma(q_{T {\rm max}})/\sigma$,
where $\sigma(q_{T {\rm max}})$ is the contribution to the total cross section
$\sigma$ from the transverse-momentum region where $q_T \leq q_{T {\rm max}}$.
The cross section $\sigma(q_{T {\rm max}})$ can be computed either at the NLO
or at NLL+LO accuracy (i.e. by integration of the NLL+LO spectrum over $q_T$).
Provided $q_{T {\rm max}}$ is not small, we find that the NLO and NLL+LO values
of $\sigma(q_{T {\rm max}})$ are very close and that $\sigma(q_{T {\rm max}})$
constitutes a large fraction of $\sigma$.
For example, the NLO (NLL+LO) value of the cross section ratio is
$\sigma(q_{T {\rm max}})/\sigma=0.86 \,(0.87)$ at $q_{T {\rm max}}=20$~GeV
and $\sigma(q_{T {\rm max}})/\sigma=0.93 \,(0.94)$ at $q_{T {\rm max}}=30$~GeV.
The resummation scale uncertainty of $\sigma(q_{T {\rm max}})$ 
at NLL+LO accuracy is
about $\pm 2$\% at $q_{T {\rm max}}=20$~GeV, and it is below the 1\% level
in the region where $q_{T {\rm max}} \gtap 30$~GeV.

In analogy with our presentation of the fixed-order results 
in Sect.~\ref{sec:fo}, we consider the resummed results and 
introduce a corresponding fractional difference 
$(X - {\rm theory})/{\rm theory}$. Now, at variance with Eq.~(\ref{fracnlo}),
we choose the NLL+LO result at central value  
of the scales as `reference theory' and we define the following
fractional difference:
\begin{equation}
\label{fracnll}
\frac{(d\sigma/dq_T)_{X} - (d\sigma/dq_T)_{NLL+LO}(\mu_F=\mu_R=2Q=m_Z)}
{(d\sigma/dq_T)_{NLL+LO}(\mu_F=\mu_R=2Q=m_Z)} \; \; ,
\end{equation}
where the label $X$ refers to either the experimental data
or the NLL+LO result at various values of the scales.

\begin{figure}[htb]
\begin{center}
\includegraphics[width=.9\textwidth]{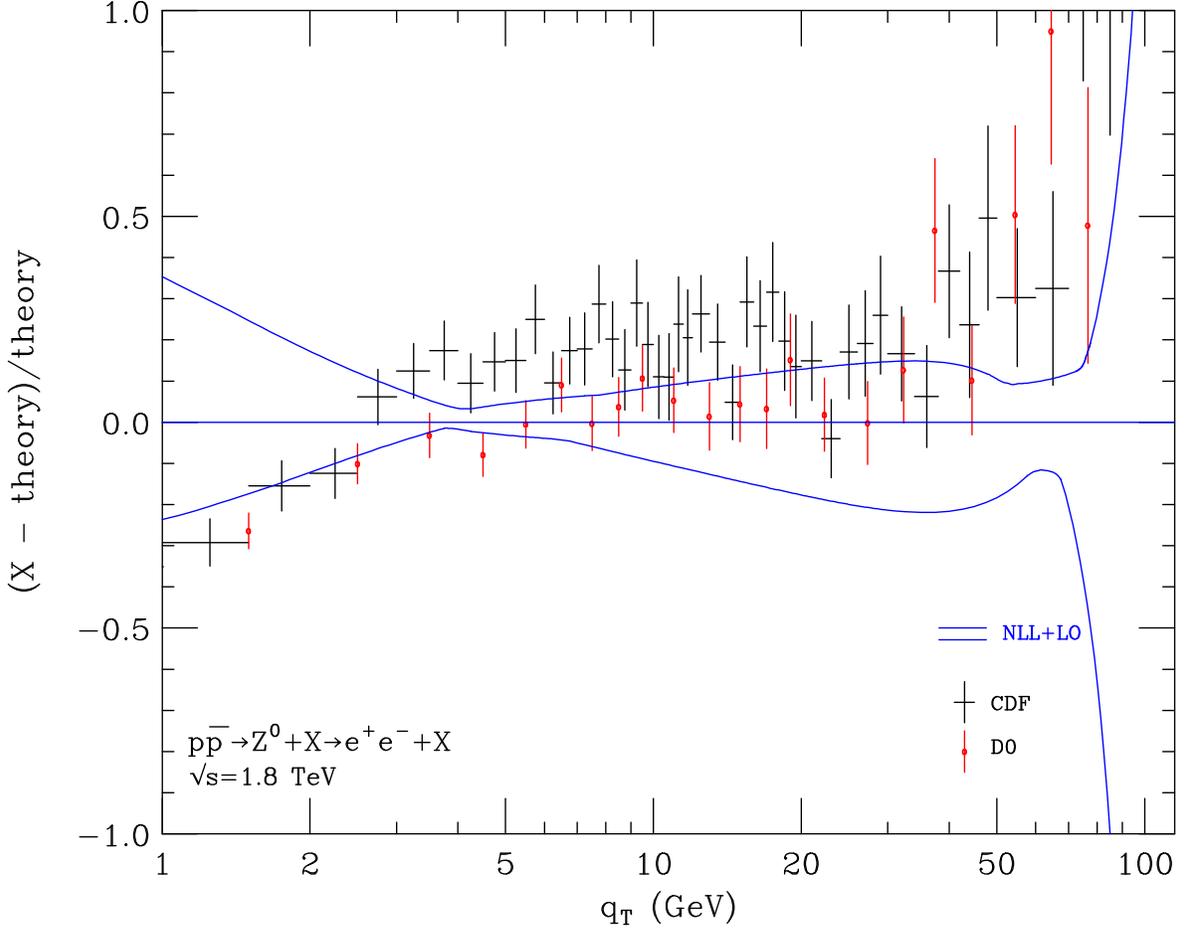}
\caption{
\it Fractional difference of Tevatron data with respect to the NLL+LO prediction at $\mu_F=\mu_R=2Q=m_Z$ (see Eq.~(\ref{fracnll})). The band represents the combined effect of varying the scales as described in the text.
}
\label{fig:fracresdata}
\end{center}
\end{figure}

This fractional difference is shown in Fig.~\ref{fig:fracresdata}.
The band enclosed by the solid lines corresponds to
our computation of the scale uncertainty. It includes the combined effect 
from varying $\mu_F, \mu_R$ and $Q$ as previously described.
We find that the resummation scale uncertainty
is larger than the factorization/renormalization scale uncertainty
at (almost) all values of $q_T$.
We comment on the results in Fig.~\ref{fig:fracresdata} by considering various
regions of transverse momenta in turn.

We first consider the large-$q_T$ region.
At $q_T \sim 40$~GeV, the scale uncertainty of the NLL+LO result is 
of about $\pm 20$\%.
At high $q_T$, the scale uncertainty is definitely larger
and quickly increases as $q_T$ increases. 
The decrease of the scale uncertainty
in the region around $q_T \sim 60$~GeV
has no physical interpretation: it has to be regarded as an accidental fact
rather than a reduction of the theoretical uncertainty.
Comparing the uncertainty of the NLL+LO calculation with that of
the fixed-order calculations (see Sect.~\ref{sec:fo}), we thus conclude that,  
in the large-$q_T$ region,
the NLL+LO predictions are less accurate than the NLO predictions.

\begin{figure}[htb]
\begin{center}
\includegraphics[width=.8\textwidth]{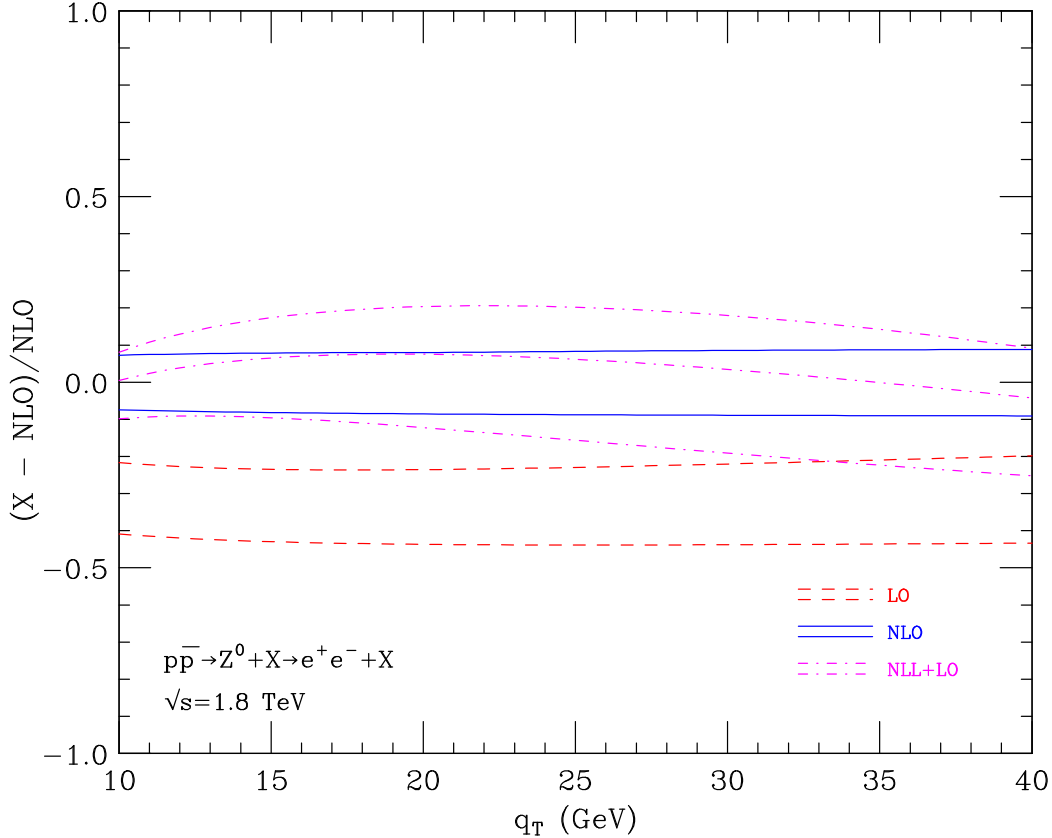}
\caption{
\it Fractional difference of LO (dashed), NLO (solid) and NLL+LO (dot dashed) predictions with respect to the NLO result at $\mu_F=\mu_R=m_Z$ (see Eq.~(\ref{fracnlo})).
}
\label{fig:fracmediumqt}
\end{center}
\end{figure}

In the region of intermediate values of $q_T$, the scale uncertainty of the 
NLL+LO result is moderate (see Fig.~\ref{fig:fracresdata}). In this region it is
appropriate to perform a more detailed comparison between the NLL+LO calculation
and the fixed-order calculations.
We consider 
Eq.~(\ref{fracnlo}), which uses 
the NLO central value as `reference theory', and we compute the fractional
difference of the NLL+LO calculation.
The results are presented in Fig.~\ref{fig:fracmediumqt}, which shows 
the NLL+LO calculation at central value  
of the scales (central dot-dashed line) and the corresponding scale uncertainty
(upper and lower dot-dashed lines). The NLO (solid lines) and LO (dashed lines)
bands in Fig.~\ref{fig:fracmediumqt} are exactly the same bands as those in 
Fig.~\ref{fig:rforun1}.
In the region where $10~{\rm GeV} \ltap  q_T \ltap 40~{\rm GeV}$, 
the NLL+LO and NLO
central values are quite close: their difference is, at most, of about
8\%.
Decreasing the value of $q_T$, the scale uncertainty at NLL+LO decreases from
about $\pm 20$\% ($q_T \sim 40$~GeV), 
to about $\pm 15$\% ($q_T \sim 20$~GeV)
and $\pm 10$\% ($q_T \sim 10$~GeV).
We also recall our conclusions (see Sect.~\ref{sec:fo})
about the uncertainty of the fixed-order perturbative expansion: the 
perturbative uncertainty of the NLO predictions is at the level of about
$\pm 20$\% in the region where $20~{\rm GeV} \ltap  q_T \ltap 40~{\rm GeV}$
(see the upper line of the LO band), and it does not decrease at smaller values
of $q_T$. We conclude that, at intermediate values of $q_T$,
the NLL+LO and NLO results are fully consistent 
and have a comparable perturbative
uncertainty. The NLL+LO calculation provides us with QCD predictions 
that can be extended to smaller values of $q_T$ ($q_T \ltap 20~{\rm GeV}$) 
with a controllable and relatively-small perturbative uncertainty.

The bulk of the production cross section is contained in the small-$q_T$
region. Considering the region above the peak of the $q_T$ distribution
($2~{\rm GeV} \ltap  q_T \ltap 20~{\rm GeV}$), the scale uncertainty
of the NLL+LO calculation is 
below the level of 
about $\pm 15$\%
(Fig.~\ref{fig:fracresdata}). The size of the scale uncertainty increases
in the region below the peak. The effect of the scale variations is larger
in the region below the peak since the shape of the $q_T$ distribution
is much steeper in this region
(see Figs.~\ref{fig:nllnlorun1} and \ref{fig:nllnloadep}).
Note also that this region is expected to be most sensitive to NP effects.
The sizeable reduction of the scale uncertainty in the interval 
$q_T \sim 3$--7~GeV can be accidental and, thus, it can underestimate
the perturbative uncertainty of the NLL+LO result.

The detailed comparison in Fig.~\ref{fig:fracresdata} shows that the
experimental data are consistent with the NLL+LO predictions in the 
small-$q_T$ region. As in the case of Fig.~\ref{fig:rforun1}, we also 
recall that part of the differences between data and theory have a systematic
component due to the luminosity uncertainties
and to the values of the total cross sections.
In particular, the central value of $d\sigma/dq_T$ at NLL+LO accuracy
corresponds to the NLO values, $\sigma=226$~pb, of the total cross section;
this value is about 9\%
smaller than the CDF value, and it is about 2\% larger than the D0 value.
These differences between the total cross sections are consistent with 
the fact that the NLL+LO result tends to agree better with the D0 data
than with the CDF data.

The quantitative predictions presented up to now are obtained in a
purely perturbative framework. It is known (see e.g. Ref.~\cite{Collins:va} 
and references therein) that the transverse-momentum distribution is affected
by NP effects, which become important
as $q_T$ becomes small.
A customary way of modelling these effects
is to introduce an NP transverse-momentum smearing
of the distribution. In the case of resummed calculations in impact parameter
space, the NP smearing is implemented by multiplying the $b$-space
perturbative form factor by an NP form factor. 
Different procedures to relate the two form factors and 
several different parametrizations of the NP form factor are available 
in the literature
\cite{Davies:1984sp, Ladinsky:1993zn, Qiu:2000ga};
the corresponding NP parameters are obtained by global fits to DY data
\cite{Qiu:2000ga, Landry:2002ix, Kulesza:2003wi}.

\begin{figure}[htb]
\begin{center}
\includegraphics[width=.8\textwidth]{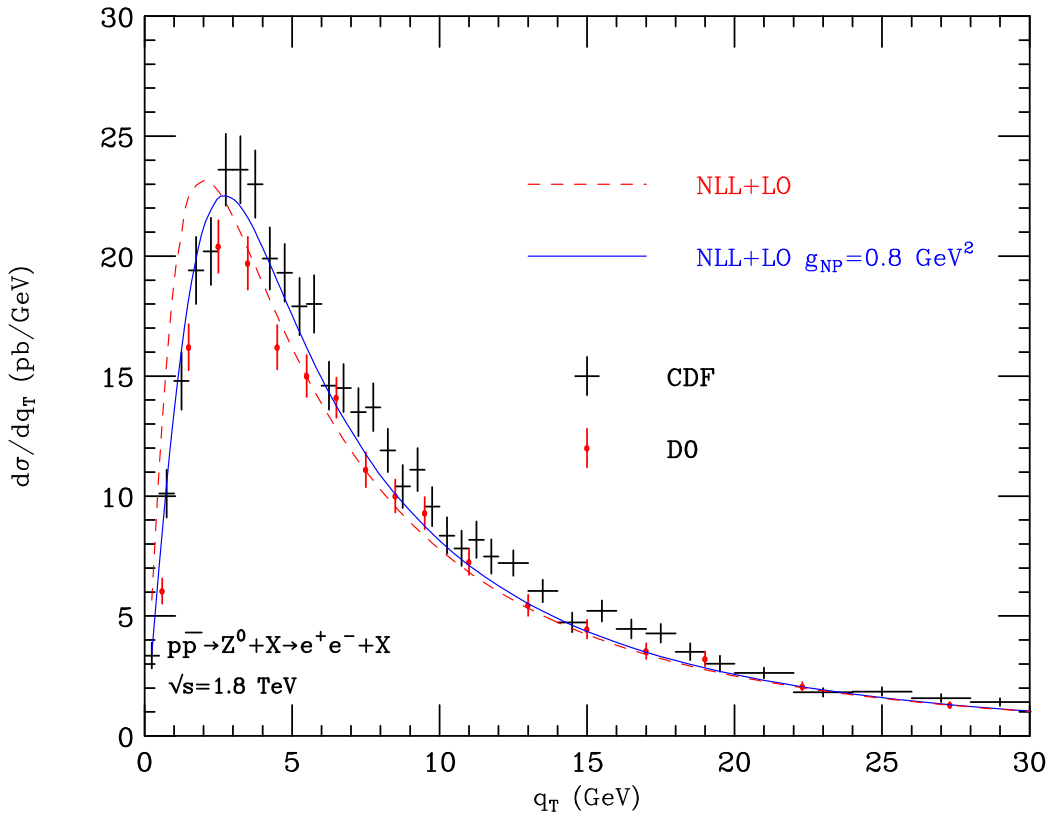}
\caption{
\it The NLL+LO spectrum at the Tevatron Run I supplemented with the NP form factor of Eq.~(\ref{NPfactor}).
}
\label{fig:Rcdfd0RI-npmio}
\end{center}
\end{figure}

A detailed study of the NP effects is beyond the aim of this work.
We limit ourselves to illustrate the possible impact of NP effects on our
resummed calculation. To this purpose, we simply multiply
the $b$-space resummation factor ${\cal W}_{N}^{V}(b,M)$ (see Eqs.~(\ref{resum}) and
(\ref{wtilde})) by 
a NP factor, $S_{NP}$, which includes a gaussian smearing of the form
\begin{equation}
\label{NPfactor}
S_{NP}=\exp\{-g_{NP}\, b^2\}\, .
\end{equation}
In Fig.~\ref{fig:Rcdfd0RI-npmio} we show the NLL+LO distribution 
at central values of the scales with (solid) and without (dashed) the 
inclusion of the NP factor.
The numerical value of the NP coefficient $g_{NP}$ is taken
to be $g_{NP}=0.8~{\rm GeV}^2$ \cite{Kulesza:2002rh}.
Using this values of $g_{NP}$, the NP effects reduce (increase) the
perturbative distribution where $q_T \ltap 3$~GeV ($q_T \gtap 3$~GeV).
For instance, 
the NP correction is about $-10$\% at $q_T \sim 2$~GeV, about
$+8$\% at $q_T \sim 5$~GeV, and it is smaller than $+4$\% where 
$q_T \gtap 10$~GeV.
As expected, the effect of the NP form factor is to make the distribution harder,
thus 
improving the agreement 
with the experimental data at very small values of $q_T$.
On the basis of the results in Ref.~\cite{Bozzi:2005wk},
we also expect that such NP effect is qualitatively similar
to that produced by the inclusion of higher-order logarithmic contributions.
This expectation is consistent with the fact 
that the quantitative impact of the NP form factor is within 
the perturbative uncertainty of the NLL+LO result 
(see Figs.~\ref{fig:nllnloadep} and \ref{fig:fracresdata}).
A detailed comparison of Tevatron data with a full NNLL+NLO calculation,
including non-perturbative effects,
is left to future work.

\begin{figure}[htb]
\begin{center}
\includegraphics[width=.9\textwidth]{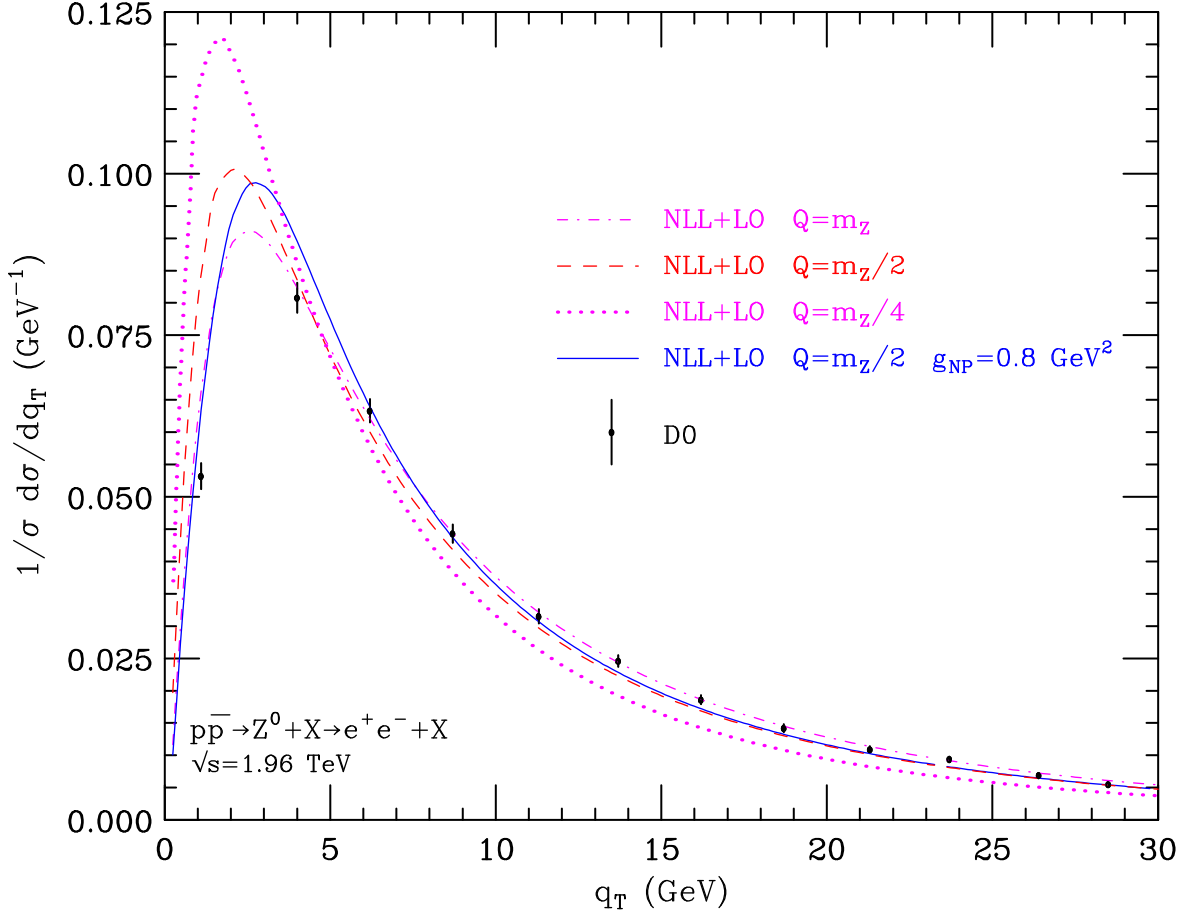}
\caption{
\it The NLL+LO $q_T$ spectrum at the Tevatron Run II,
for different values of the resummation scale $Q$. The effect of including
a NP smearing as in Eq.~(\ref{NPfactor}) is also shown.
}
\label{fig:d0RIInp}
\end{center}
\end{figure}

We have repeated our NLL+LO study by considering the
normalized $q_T$ distribution, $\frac{1}{\sigma} \frac{d\sigma}{dq_T}$,
at the Tevatron Run~II. The main features of the results are definitely very
similar to those at Run~I. The NLL+LO results at small and intermediate values 
of $q_T$ are briefly summarized in Fig.~\ref{fig:d0RIInp}.
We show the NLL+LO distribution 
at central values of the scales with (solid) and without (dashed) the 
inclusion of the NP form factor (we use the same values of $g_{NP}$ as in 
Fig.~\ref{fig:Rcdfd0RI-npmio}). Since the perturbative uncertainty is dominated
by the effect of varying the resummation scale $Q$, we fix 
$\mu_R=\mu_F=m_Z$ and we also show the NLL+LO results with $Q=m_Z$ (dotdashed)
and $Q=m_Z/4$ (dotted). We observe that the D0 data are consistent with the 
NLL+LO predictions. We also observe that the experimental errors are 
smaller than the uncertainty of the NLL+LO results, thus demanding more accurate
perturbative predictions.

\section{Summary}
\label{sec:summa}

We have considered the $q_T$ cross section of DY $e^+e^-$ pairs 
from the decay of $Z$ bosons produced in $p{\bar p}$ collisions at Tevatron
energies.
 
We have presented fixed-order QCD predictions up to NLO, including an estimate
of the corresponding perturbative uncertainty.
In the region of large and intermediate values of $q_T$, the CDF and D0 data
at Tevatron Run~I show an overall agreement with the NLO results. Some
deviations from the NLO predictions
are observed in the Run~II D0 data at moderate values of $q_T$.
In the region of small values of $q_T$, the comparison between the LO and NLO
results shows the onset of instabilities of the order-by-order perturbative 
expansion. As $q_T$ decreases toward very small values, the fixed-order 
central values
definitely disagree with the Tevatron data. 

As is well known, in the small-$q_T$ region, 
there are large logarithmic contributions that spoil the reliability of the
fixed-order perturbative expansion and that 
need be resummed to all orders. 
We have presented a first application of the resummation
formalism of Refs.~\cite{Bozzi:2005wk} to 
$Z$ boson production.
The formalism combines small-$q_T$ resummation at a given logarithmic accuracy 
with the fixed-order calculations. It implements a unitarity constraint
that guarantees that the integral over $q_T$ of the differential cross section
coincides with
the total cross section at the corresponding fixed-order accuracy.
The formalism includes the explicit dependence on the 
factorization and renormalization scales, 
analogously to the customary scale dependence in fixed-order calculations.
It also introduces an auxiliary resummation scale, whose dependence can
be exploited to estimate the effect of uncalculated higher-order logarithmic
contributions. Owing to these features, the resummation formalism 
extends the applicability of QCD perturbation theory to the small-$q_T$  
region, with a controllable perturbative accuracy at small and intermediate
values of $q_T$.

We have presented the results of the resummed calculation at NLL+LO accuracy,
and we have performed a detailed study of the scale dependence to estimate 
the corresponding perturbative uncertainty. In the region of
intermediate values of $q_T$,
the NLL+LO and NLO results are fully consistent 
and have a comparable perturbative
uncertainty. In the small-$q_T$ region, the Tevatron data are consistent with
the NLL+LO predictions, 
which are not supplemented with additional   
non-perturbative contributions.
The perturbative uncertainty of the NLL+LO results,
which is dominated by missing higher-order logarithmic contributions, is
relatively large in comparison with the precision of the experimental data.
This uncertainty is likely to be comparable with the effect of
non-perturbative contributions.
On the basis of the results on $q_T$ spectrum
of the Standard Model Higgs boson 
\cite{Bozzi:2003jy,Bozzi:2005wk,Bozzi:2007pn},
we expect a reduction of the perturbative uncertainty
once the complete NNLL+NLO calculation for the DY process is available.

\end{document}